\documentclass{ws-rv9x6}
\usepackage{ws-rv-thm}   % comment this line when `amsthm / theorem / ntheorem` package is used
\usepackage{ws-rv-van}   % numbered citation & references (default)
\usepackage{hyperref}
\usepackage{listings}
\usepackage{dcolumn}
\usepackage{bm}
\makeindex
\graphicspath {{./figures/}}
\makeatletter
\def\input@path{{./figures/}}
\makeatother

\newcolumntype{t}[1]{D{.}{.}{#1}}
\definecolor{dkgreen}{rgb}{0,0.6,0}
\definecolor{gray}{rgb}{0.5,0.5,0.5}
\definecolor{mauve}{rgb}{0.58,0,0.82}
\begin{document}
\chapter{Monte Carlo methods for massively parallel computers}

\author[M. Weigel]{Martin Weigel}

\address{Applied Mathematics Research Centre, Coventry University\\Coventry, CV1 5FB, United Kingdom\\
  martin.weigel@complexity-coventry.org}

\begin{abstract}

  Applications that require substantial computational resources today cannot avoid
  the use of heavily parallel machines. Embracing the opportunities of parallel
  computing and especially the possibilities provided by a new generation of
  massively parallel accelerator devices such as GPUs, Intel's Xeon Phi or even FPGAs
  enables applications and studies that are inaccessible to serial programs. Here we
  outline the opportunities and challenges of massively parallel computing for Monte
  Carlo simulations in statistical physics, with a focus on the simulation of systems
  exhibiting phase transitions and critical phenomena. This covers a range of
  canonical ensemble Markov chain techniques as well as generalized ensembles such as
  multicanonical simulations and population annealing. While the examples discussed
  are for simulations of spin systems, many of the methods are more general and
  moderate modifications allow them to be applied to other lattice and off-lattice
  problems including polymers and particle systems. We discuss important algorithmic
  requirements for such highly parallel simulations, such as the challenges of
  random-number generation for such cases, and outline a number of general design
  principles for parallel Monte Carlo codes to perform well.
 
\end{abstract}
\body

\tableofcontents

\section{Introduction}
\label{sec:intro}

The explosive development of computer technology over the past 40 years or so has not
only led to pervasive changes of the industrial world and to the way we communicate,
learn, work, and entertain ourselves, but it has also enabled an impressive success
story of computational sciences \cite{gramelsberger:11}. In condensed matter and
statistical physics, numerical methods such as classical and quantum molecular
dynamics \cite{rapaport:04}, density functional theory \cite{sholl:09} and Monte
Carlo simulations \cite{binder:book2} were initially developed in the late 1950s and
early 1960s when the first digital computers became available. Before that, the tool
set of theoretical physics was restricted to exact solutions for sufficiently
simplified systems, mean-field type theories neglecting fluctuations, and
perturbative methods such as the $\epsilon$ expansion and high-temperature
series. Due to the limited computational power available, numerical techniques were
not yet quite competitive, and some researchers considered them as inferior crutches
for people allegedly lacking the brilliance for analytical work. It is very rare
indeed that one hears such opinions expressed today, and simulations are now firmly
established as an indispensable scientific method, a third pillar supporting the
building of science besides those of experiment and analytical theory
\cite{binder:book2}.

This success is the result of two parallel developments: the enormous increase of
computational power by a factor of at least $10^7$ since the first digital computers
appeared \cite{moore:65}, but no less the development of ever more sophisticated
simulation and other computational methods enabling calculations that were unfeasible
with simpler techniques. For simulations in statistical physics the focus of advanced
methods has been the study of systems experiencing phase transitions and critical
phenomena as well as other effects of complexity such as exotic phases with slow
relaxation. Here, one should name cluster updates \cite{swendsen-wang:87a,wolff:89a}
\index{cluster~algorithm} that are effective in beating critical slowing down close
to continuous phase transitions, multicanonical simulations \cite{berg:92b,wang:01a}
\index{multicanonical~simulation} that allow to sample the suppressed co-existence
region in systems undergoing first-order \index{phase~transition} phase transitions,
and exchange Monte Carlo \cite{geyer:91,hukushima:96a} \index{parallel~tempering}
\index{exchange~Monte Carlo|see{parallel~tempering}} that is currently the workhorse
for simulations of systems with complex free energy landscapes
\index{complex~free-energy~landscape} such as spin glasses, but also methods of data
analysis such as histogram reweighting \index{histogram~reweighting} and advanced
methods of error analysis \cite{efron:book,weigel:10}. Only combining both strengths,
i.e., using advanced algorithms on sufficiently powerful hardware enables computer
simulation studies to achieve the level of detail and precision required
today.\footnote{Consider, for instance, the problem of simulating the two-dimensional
  Ising model. The Metropolis algorithm has a dynamical critical exponent of
  $z = 2.17(1)$ \cite{ito:93}, while a recent estimate for the exponent of the
  Swendsen-Wang algorithm is $z = 0.14(1)$ \cite{deng:07a}. Assuming scaling
  amplitudes of approximately one in the law $\tau \sim L^z$ of the autocorrelation
  times, this results in an algorithmic speedup of $2\times 10^7$ for a realistic
  system size $L=4096$, well comparable to the total increase in computational power
  in the past 40 years.}

On the computational side, the high-performance computing (HPC) setups available
today are highly parallel in nature, and no further significant increase of serial
execution speeds of silicon based computing can be expected \cite{asanovic:06}. Some
of the best performance results, especially in terms of FLOPs per Watt, are now
achieved by parallel accelerator devices such as GPUs, Intel's \index{Xeon~Phi} Xeon
Phi and \index{FPGA} FPGAs. For computational scientists one of the most pressing
current challenges is hence the efficient implementation of existing algorithms on
such massively parallel hardware, but also, if possible, the design of new algorithms
particularly well suited for highly parallel computing. The purpose of the present
chapter is to provide some guidance for the practitioners of Monte Carlo methods
particularly in statistical physics as we are moving further into the era of parallel
computing. The focus is on simulations of spin models on graphics processing units
and using a wide range of algorithms, but we will see that many of the general
concepts and design principles are also useful for simulations of different lattice
and continuum models and for different hardware such as MPI clusters and Intel's Xeon
Phi family of co-processors.

The rest of the chapter is organized as follows: Section \ref{sec:parallel} discusses
the necessary background in parallel computing, including some standard algorithmic
patterns for efficient parallelism, as well as the relevant parallel hardware
including, in particular, an outline of the most important architectural features of
graphics processing units. In Sec.~\ref{sec:canonical} we discuss implementations of
standard local-update algorithms, such as the Metropolis and heatbath updates for the
example of discrete spin models. While these can be realized rather straightforwardly
using domain decompositions, we next turn to non-local cluster algorithms that are
more difficult to parallelize efficiently as they operate on clusters that percolate
at the critical point. Finally, we discuss the specific problems of simulating
systems with continuous variables on GPU, that arise due to the performance penalty
paid for double precision floating point arithmetics on such devices. Section
\ref{sec:RNG} is devoted to a discussion of random-number generation in highly
parallel environments, where the availability of a large number of uncorrelated
streams of random numbers is required. In Sec.~\ref{sec:generalized} we turn to
parallel implementations of generalized-ensemble simulations, discussing the cases of
parallel tempering, multicanonical and Wang-Landau simulations, and a variant of
so-called sequential (non Markov-chain) Monte Carlo known as population annealing
\cite{iba:01,hukushima:03} that has recently attracted some attention
\cite{machta:10a}. Section~\ref{sec:disordered} is devoted to a discussion of the
specific challenges and opportunities that are held by parallel machines for the
treatment of systems with random disorder, where the necessary quenched average
provides the possibility for embarrassingly parallel implementations. Finally,
Sec.~\ref{sec:conclusions} contains our conclusions.

\section{Parallel computing}
\label{sec:parallel}

\begin{figure}[tb]
  \centering
  \includegraphics[clip=true,keepaspectratio=true,width=0.49\columnwidth]{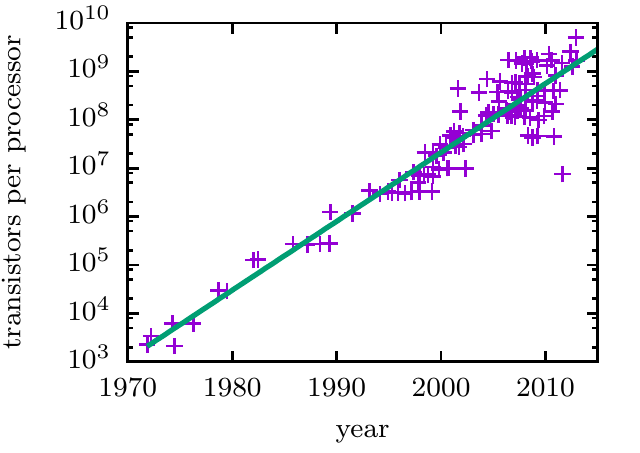}
  \includegraphics[clip=true,keepaspectratio=true,width=0.49\columnwidth]{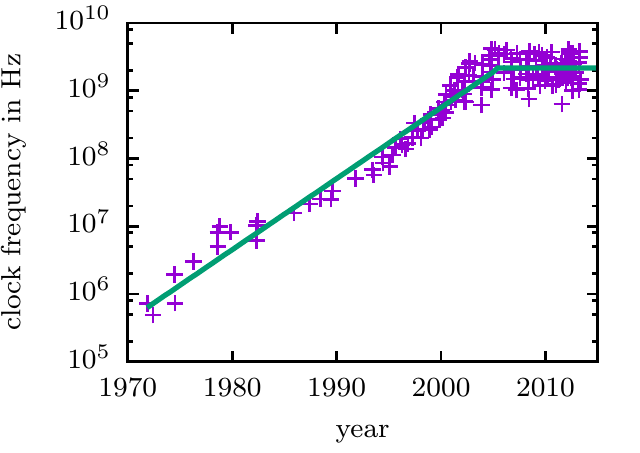}
  \caption
  {Left: number of transistors on commodity processors as a function of their release
    year. The line shows a pure exponential fit to the data, illustrating the
    validity of Moore's law \cite{moore:65}. This fit implies a doubling about every
    $2.1$ years. Right: clock frequencies of the same CPUs, showing the leveling off
    of (approximately exponential) frequency increases around 2003. The data are
    adapted from Ref.~[\refcite{mccool:12}].
    \label{fig:moores_law}
  }
\end{figure}

Gordon Moore's prediction \index{Moore's~law} from 1965 of a doubling of the number
of transistors every two years has been a surprisingly accurate description of the
development of integrated circuits over the last four decades \cite{moore:65}. Figure
\ref{fig:moores_law} (left) illustrates this for the case of Intel CPUs showing an
increase by 7 orders of magnitude from about 1000 transistors in 1970 up to almost
$10^{10}$ transistors today. Although there are clearly physical limits to this
development, these are not yet seen in processors today and hence the development can
be expected to continue unabated for a while. Another characteristic of computer
processors, however, the clock \index{clock~frequency} frequency, which for decades
showed an equally dynamic increase, started to settle down at around 3 GHz in 2003,
see the data of historic CPU clock frequencies shown in the right panel of
Fig.~\ref{fig:moores_law}. It turns out that an increase of clock frequencies beyond
a few GHz is not practically feasible for commodity hardware, mostly because the
electrical power consumption increases dramatically with the frequency\footnote{The
  dynamic power consumption is in fact given by $P \propto V^2f$, where $V$ is the
  operating voltage and $f$ the frequency \cite{mccool:12}. Since, however, the
  highest operating frequency $f$ is itself proportional to the voltage $V$, in total
  $P\propto f^3$.} and there is a natural limit to the maximal power density that can
be dissipated with conventional cooling methods. Due to the leveling off of clock
frequencies, but also through further effects including limitations in exploiting
instruction-level parallelism and the fact that speeds of memory technologies have
not developed as dynamically as those of processing units, recently there has been
hardly any relevant improvement in the speed of serial programs on standard
processors. For decades, scientific and application programmers have been in the
comfortable situation that the same serial program could be run on a series of
generations of processors and its speed would improve exponentially according to
Moore's law, thus allowing scientists to study ever larger system sizes and all users
to process bigger data sets with the same codes as time progressed. This development
has now come to an end.

The way that Moore's law continues to hold while serial performance has reached a
limit is, of course, through the introduction of more and more parallel
cores. Typical CPUs are now multi-core \index{CPU!multi-core} with up to a few ten
cores, accelerator devices such as Intel's MIC (Xeon Phi) architecture are many-core
with dozens to hundreds of cores and GPUs offer several thousand cores in one
device. In short, all computers are now parallel, from multi-core processors in
mobile phones up to the top machines in the TOP500 list of supercomputers with
millions of cores \cite{top500}. Consequently, programs that make efficient use of
present-day machines must be parallel codes. While modern compilers have some
capabilities of automatic parallelization, these are quite limited and they will
typically not generate code that scales well on machines with different numbers of
cores. Apart from any shortcomings in the compilers themselves, this is mainly a
consequence of the serial nature of the prevalent programming languages themselves
which produce what could be called \index{implicit~serialism} implicit serialism: if
a task requires several different steps, these must be written in sequence in a
serial language, for example as a list of function calls; a compiler cannot always
decide whether such steps are independent and hence could be performed in parallel,
or whether some of them have side effects that influence the other steps. Similarly,
loop constructs cannot be parallelized automatically when they contain pointer
arithmetic or possibilities for overlapping index ranges. Other examples occur for
sums or more general reductions \index{algorithmic~pattern!reduction} involving
floating-point numbers: due to the limited accuracy such operations are not
commutative, and a reordering of the sum will lead to a (most often slightly)
different result, typically disabling automatic parallelization to ensure
consistency, although the ensuing rounding differences might be perfectly acceptable
in a given application. Programs in serial languages and the tools to process them
contain many such serial assumptions. As today programming is in fact parallel
programming, it is crucial to get rid of the implicit assumption of seriality in
thinking about algorithms and augment if not replace the well-known serial
algorithmic building blocks (such as \index{algorithmic~pattern!iteration} iteration,
\index{algorithmic~pattern!recursion} recursion etc.) by parallel ones (such as
\index{algorithmic~pattern!fork-join} fork-join or
\index{algorithmic~pattern!scatter} scatter) \cite{mccool:12}.

A variety of parallel programming languages or language extensions have been proposed
to support this transition. MPI \index{MPI} is the de facto standard for distributed
memory machines such as cluster computers \cite{mpi}. OpenMP \index{OpenMP} is very
popular for shared memory machines such as single nodes with (one or several)
multi-core processors, especially for applications in HPC. Its explicit
representation of threads allows fine control in specific situations, but a single
code will typically not scale well across many different types of hardware ranging
from embedded systems to supercomputers. This goal is more easily achieved using
language extensions such as Threading Building Blocks (TBB), Array Building Blocks
(ArBB), or Cilk Plus \cite{mccool:12}. Finally, frameworks for accelerator devices,
in particular GPUs, include the vendor-specific Nvidia CUDA \index{CUDA} toolkit as
well as OpenCL \index{OpenCL} \cite{kirk:10}. A detailed discussion of different
programming models is clearly outside of the scope of the present chapter and the
interested reader is referred to the literature, for instance the excellent
Ref.~[\refcite{mccool:12}]. Although there are many differences between these
approaches, a general goal of any such framework must be the creation of scalable
code that is able to run efficiently on any amount of parallel hardware and in a
performance portable manner, promising decent efficiency also on the next generation
of machines. Other desirable features are \index{composability} {\em
  composability\/}, i.e., the possibility to use all language features together in
the same code, as well as \index{determinism} {\em determinism\/}, i.e., a guarantee
that each invocation of the program leads to identical results. The latter feature is
very useful for testing and debugging purposes and it is natural for serial codes,
but in some cases it might be difficult to achieve (and detrimental to performance)
in parallel programs where the scheduling of individual threads is typically outside
of the programmer's control.

The two basic strategies for parallelization are \index{parallelism!data} data
parallelism and \index{parallelism!functional} functional decomposition. While the
latter can create a limited amount of parallel work, it is clear that only data
parallelism, for example in the form of \index{domain~decomposition} domain
decomposition, creates a number of tasks that scales with the size of the
problem. This will also be most often the type of parallelism encountered in
simulation codes where different parts of the system are assigned to different
threads. Functional parallelism, on the other hand, could occur in the present
context for complex simulations on heterogeneous machines with accelerators, where
only parts of the calculations (for example force-field evaluations) are offloaded to
the accelerators and the remaining computations are run on the host machine
\cite{hwu:11}. It is sometimes also useful to distinguish \index{parallelism!regular}
regular and \index{parallelism!irregular} irregular parallelism, where the regular
kind has predictable and regular dependencies such as for the case of matrix
multiplication, while irregular parallelism could occur in a parallel evaluation of a
search tree through a branch-and-bound \index{branch-and-bound} strategy, such that
some branches and hence parallel tasks are terminated early through the bounding
step. As we will see below in Sec.~\ref{sec:cluster}, irregular parallelism occurs in
the tree-based methods for cluster updates of spin models. In terms of mechanisms,
parallel computation can be through threads or through vector parallelism. Threads
have a separate control flow, while vector calculations apply the same instructions
to a vector of data elements in parallel. Clearly,\index{parallelism!thread} thread
parallelism can emulate \index{parallelism!vector} vector parallelism. As we shall
see below for the case of GPUs, vector parallelism can also emulate thread
parallelism through the masking out of operations for some of the data elements
(lanes) of the vector (Sec.~\ref{sec:hardware}). Such vector parallelism hence
creates pseudo-threads sometimes called \index{fiber} fibers.

\subsection{Performance and scaling}
\label{sec:scaling}

Ideal parallel programs will run efficiently on a wide range of hardware with
possibly very different numbers of cores. Code that achieves such performance
portability cannot explicitly depend on the features of particular hardware, for
example its specific memory hierarchy. This approach can achieve good but generally
not optimal performance. In HPC applications, on the other hand, it is often
admissible to be somewhat more specific to a class of hardware and thus use a larger
fraction of the available peak performance. In most cases, however, taking the right
general design decisions will contribute significantly more to achieving good
performance than machine-specific optimizations.

\begin{figure}[tb]
  \centering
  \includegraphics[clip=true,keepaspectratio=true,width=0.49\columnwidth]{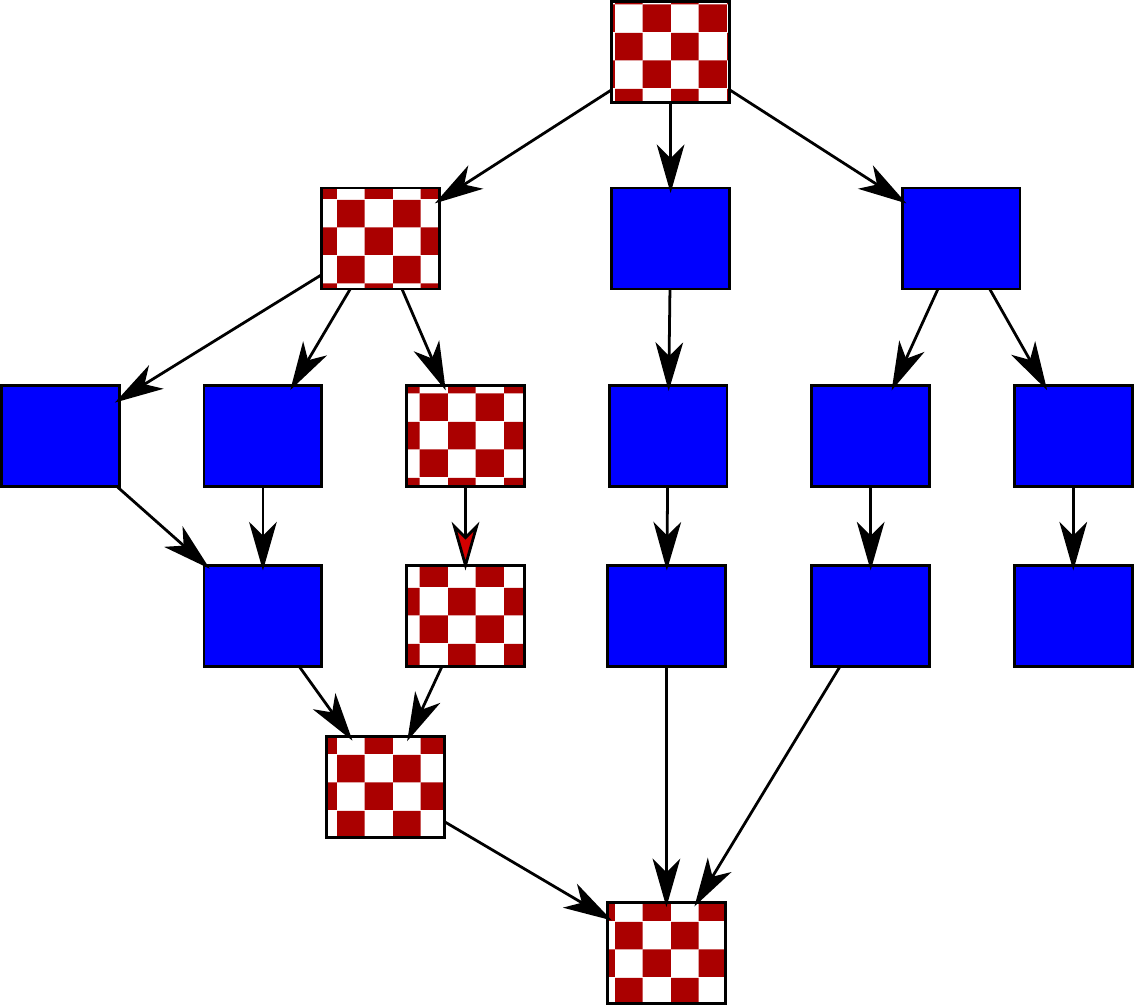}
  \caption
  {Example of the steps in a computation represented as an acyclic, directed graph of
    task dependencies, assuming the same duration for each task. The total number of
    tasks, 17 in the example, corresponds to the work in the computation, while the
    longest path from start to end configuration, 6 in the present example, is the
    span of the algorithm (hatched squares).}
  \label{fig:workspan}
\end{figure}

The limiting factors for parallel performance are (data and control)
\index{dependency} dependencies between tasks and the communication required between
them. A good framework for estimating performance is the {\em work-span model\/}
\index{work-span~model} \cite{blelloch:96}: if one represents the necessary steps of
a calculation in an acyclic, directed graph of tasks with edges encoding the
dependencies, the span is the time it takes to perform the longest chain of
instructions that cannot be parallelized (possibly including the effects of
synchronization and communication overheads). This is illustrated in
Fig.~\ref{fig:workspan}. The span limits the possible parallel speedup and,
consequently, reducing the span is arguably the most important step towards an
efficient parallel program. This could be through the removal of implicit serialism,
i.e., getting rid of assumed dependencies that are unnecessary, or through more
profound reorganizations of calculations. Apart from the goal of reducing an
algorithm's span, the two most profound considerations when performance optimizing a
parallel application are data locality and parallel slack \cite{kirk:10,mccool:12}:
\begin{itemize}
\item {\bf Data locality}:
  \index{data~locality|see{memory~locality}}
  \index{memory!locality|(}%
  memory accesses that are closer together in time and space
  are cheaper. Fetching close-by memory locations together makes the best use of bus
  bandwidth, reusing close-by data makes the best use of caches. As a general rule,
  memories that are further remote from the compute units have slower connections and
  higher latencies, so memory transactions should be restricted to local memories as
  much as possible. This might involve choosing chunk sizes that fit into cache, the
  reorganization of the memory layout to ensure that subsequent accesses are to
  nearby locations, or padding to achieve the required memory alignment. As memory
  transactions are so expensive, it is important to ensure sufficient arithmetic
  intensity of computations --- sometimes it is cheaper to recompute intermediate
  results than to read them from memory.%
  \index{memory!locality|)}
\item {\bf Parallel slack}:
  \index{parallel~slack|(}%
  providing more parallel tasks
  than cores are available improves efficiency. It might be tempting to break a
  problem into exactly as many threads as can be run in parallel on the available
  hardware, but this is typically not optimal. Having more threads than cores allows
  the scheduler to hide memory latencies by putting thread groups waiting for memory
  accesses into a dormant state while reactivating other thread groups that have
  received or written their data. In general, it is best to break the calculation
  into the smallest units that can still amortize the overhead of scheduling a
  thread.%
  \index{parallel~slack|)}
\end{itemize}

The main aspects of computational performance concern \index{latency} {\em
  latency\/}, i.e., the total time it takes to finish a single calculation, as well
as \index{throughput} {\em throughput\/}, i.e., the rate at which a sequence of
calculations can be performed. Increasingly, also the power consumption of a
calculation is considered as a separate performance metric \cite{green500}. Depending
on the application, the reduction of latency or the improvement of throughput might
be the main goal of optimization. The most common metric is the \index{speedup} {\em
  speedup in latency\/},
\[
  S_p = \frac{T_1 W_p}{W_1 T_p},
\]
where $T_1$ ($T_p$) is the latency and $W_1$ ($W_p)$ denotes the workload of the
problem with one worker ($p$ workers). The speedup per worker, $S_p/p$, is known as
{\em parallel efficiency\/} which indicates the return on adding an additional
worker. Clearly, the ideal efficiency is 100\% corresponding to linear speedup,
although in some unusual circumstances one finds $S_p/p > 1$ due, for example, to
cache effects. In determining parallel speedup, one should compare to the best serial
program available, even if it uses a different algorithm. The corresponding absolute
speedup arguably provides a fairer comparison than the relative speedup of running
the parallel code with just one thread. In a similar way one can also define and
analyze the speedup in throughput. An essential aspect of parallel performance theory
relates to the scaling of performance with $p$. Two important limits relate to the
cases of a fixed amount of work $W$ performed with a variable number $p$ of
processors, corresponding to \index{scaling!strong} {\em strong scaling\/}, and the
situation where the problem size and hence the amount of work are scaled proportional
to $p$, known as \index{scaling!weak} {\em weak scaling\/}. In the strong scaling
scenario with work $W$, the latency of the serial program is proportional to $W$,
$T_1 = t W$. If we assume that the work decomposes into parallelizable and
intrinsically serial parts, $W = W_\mathrm{par} + W_\mathrm{ser}$, the latency for
the parallel execution satisfies $T_p \ge t(W_\mathrm{ser} + W_\mathrm{par}/p)$ and
hence the maximal parallel speedup is limited by
\[
  S_p = \frac{T_1W}{T_pW} \le
  \frac{W_\mathrm{ser}+W_\mathrm{par}}{W_\mathrm{ser}+W_\mathrm{par}/p}.
\]
\index{Amdahl's~law|(}%
If the serial part makes up a constant fraction $f$ of the work,
$W_\mathrm{ser} = f W$ and $W_\mathrm{par}=(1-f)W$, we have
\begin{equation}
  S_p \le \frac{1}{f+(1-f)/p}
  \label{eq:amdahl}
\end{equation}
and hence the speedup is limited by $S_\infty \le 1/f$ such that, for example, an
algorithm that has 10\% intrinsically serial calculations cannot be sped up beyond a
factor of ten, no matter how many cores are available. Eq.~(\ref{eq:amdahl}) is known
as Amdahl's law \cite{amdahl:67}.
\index{Amdahl's~law|)}%

In practice, problem sizes are often scaled with the number of available cores, and
so the assumption of constant work might not be appropriate. If, instead, the
parallel work increases proportional to $p$, i.e.,
\[
W_p = W_\mathrm{ser} + p W_\mathrm{par} = f W_1 + p(1-f)W_1,
\]
where it was again assumed that the not-parallelizable work makes up a fraction $f$
of $W_1$, and we consider the work done in a fixed time budget $T$, the speedup in
latency becomes
\begin{equation}
  S_p = \frac{TW_p}{TW_1} = f + (1-f)p,
  \label{eq:gustafson}
\end{equation}
which is asymptotically proportional to $p$ as $p\to\infty$. The relation of
Eq.~(\ref{eq:gustafson}) is referred to as Gustafson-Barsis \index{Gustafson's~law}
law for weak scaling \cite{gustafson:88}.

A more fine grained analysis of parallel performance of algorithms is possible in the
work-span model \index{work-span~model} outlined above. There, the time for one
worker, $T_1 = tW$, is called the work, choosing units such that $t=1$ for
simplicity. The time $T_\infty$ for an infinite number of workers is called the
span. It corresponds to the length of the longest chain in the execution graph for an
infinite number of workers. We easily see that $S_p \le p$, so super-linear speedup
is impossible in this model. On the ideal machine with greedy scheduling, adding a
processor can never slow down the code, such that
\[
S_p = \frac{T_1}{T_p} \le \frac{T_1}{T_\infty},
\]
so the speedup is limited by the ratio of work and span. If the work consists of
perfectly parallelizable and imperfectly parallelizable parts, the latter will take
time $T_\infty$, irrespective of $p$. The former then takes time $T_1-T_\infty$ when
using one worker and, as it is perfectly sped up by additional cores, time
$(T_1-T_\infty)/p$ with $p$ workers. As at least one worker needs to be dealing with
the imperfectly parallelizable part, this provides an upper bound known as Brent's
lemma \cite{mccool:12},
\[
T_p  \le (T_1-T_\infty)/p+T_\infty,
\]
which provides a {\em lower\/} bound on the parallel speedup. From this, a good
practical estimate for $T_p$ can be derived noting that for problems suitable for
parallelization we must have $T_1 \gg T_\infty$ and hence
\[
T_p \approx T_1/p + T_\infty.
\]
Hence it is clear that the span is the fundamental limit to parallel scaling. From
Brent's lemma one derives that if for $S_\infty = T_1/T_\infty \gg p$, i.e., if the
theoretical maximal speedup is much larger that the actually available parallelism,
the speedup is approximately linear, $S_p \approx p$. Hence it is good to have
sufficient \index{parallel~slack} {\em parallel slack\/}, a standard recommendation
is $S_\infty/p \ge 8$. This is called over-decomposition.

\subsection{Parallel hardware}
\label{sec:hardware}

While a number of general design principles, most notably the concepts of data
locality and parallel slack outlined above, will contribute to good performance of
parallel programs independent of the hardware, a substantial fraction of the peak
performance can typically only be achieved with some tailoring to the hardware to be
used.

A common classification of parallel processing paradigms relates to the way that
control flow and data are combined \cite{flynn:72}: single instruction, single data
(SISD) \index{SISD} setups correspond to standard serial processing; single
instruction, multiple data (SIMD) \index{SIMD} approaches imply vector processing
with an array of functional units performing identical calculations on different data
elements; multiple instruction, multiple data (MIMD) \index{MIMD} corresponds to
separate instruction streams, each applied to their own data sets --- this is
implemented in a cluster computer. Another classification concerns memory
organization: in shared memory machines each compute element can access all data,
whereas in distributed memory setups this is not possible. Cluster machines are
examples of the latter type, where data between different nodes can only be accessed
after explicitly communicating it between them. Each node, on the other hand, will
typically feature several cores that operate a shared memory setup between them.

Parallelism occurs in current hardware at many different levels. At the scope of a
single CPU core there is instruction-level parallelism in the form of
\index{parallelism!superscalar} superscalar execution of serial instructions, through
hardware multi-threading and vector instructions in extensions
\index{parallelism!vector~extensions} such as SSE and AVX. These features are
generally hard to configure explicitly unless programs are written in assembly
language, and they will often only be activated through certain compiler
optimizations. Modern CPUs come with multiple cores \index{CPU!multi-core} and hence
can run multiple, and possibly many, threads. Such parallelism is typically only
accessible to programs that are explicitly parallel, using multi-threading language
extensions such as OpenMP, TBB, ArBB or Cilk Plus. To ensure good performance, data
locality needs to be respected, and it is hence important to understand the memory
hierarchy of multi-core CPU systems: the functional units are equipped with a
moderate number of very fast registers, and a cascade of cache memories (typically
L1, L2 and L3) translates accesses down to the main memory of the machine. In
general, bandwidths decrease and latencies increase as the hierarchical (and thus the
physical) distance of memory locations to the compute units increases. Caches are
typically organized in lines of 64 or 128 bytes, and each access to main memory
fetches a full cache line, thereby accelerating accesses to nearby memory
locations. Only \index{memory!coherence} coherent accesses therefore allow to achieve
transfer rates close to the theoretical memory bandwidths. Finally, there is also a
virtual memory \index{memory!virtual} system underneath the actual physical memory,
swapping pages of unused memory out to disk as required, and a lack of memory
locality \index{memory!locality} will lead to frequent page faults that are immensely
expensive on the timescale of the CPU clock.

\index{accelerators|(}
\index{Xeon~Phi|(}
\index{GPU|(}

A relatively recent addition to the arsenal of parallel hardware are accelerator
devices such as GPUs, Intel's MIC (many integrated core) processors, and
field-programmable gate arrays (FPGAs). GPUs and MIC devices provide a large number
of relatively simple compute cores packaged on a separate device which is used to
offload expensive calculations that are well suited for parallel execution. While
GPUs use specific programming models (see below), the Intel MIC architecture appears
to the user like a standard multi-core system with particularly many cores (currently
around 60), supporting most of the standard development tool-chain. FPGAs, on the
other hand, are integrated circuits that can be reconfigured on demand to implement
an algorithm in hardware. While traditionally, this could only be achieved by experts
in circuit development using a hardware description language, it is now possible to
use general-purpose programming languages (with suitable extensions) to configure
FPGAs, for example OpenCL \cite{moore:17}. For particular parallel applications,
FPGAs can provide higher performance at a lower power consumption than any other
parallel hardware.

\index{accelerators|)}
\index{Xeon~Phi|)}

GPUs operate at a sweet spot of parallel computing in that they provide very
substantial parallelism with the availability of several thousand parallel hardware
threads, but without requiring an expensive distributed memory machine such as a
cluster computer. As most of the implementations presented in the application part of
this chapter have been realized for GPUs, we discuss their architecture in somewhat
more detail here. Clearly, GPUs have been designed for the efficient rendering of
(mostly 3D) computer graphics, a task that involves the parallel manipulation of many
3D objects, the mapping of textures, and the simultaneous projection of a scene onto
the millions of pixels in an image frame. Driven by the large sums of money available
through the gaming industry, GPUs are hence highly optimized to perform well for
these massively parallel and very predictable tasks. For a number of generations,
their peak performances have substantially exceeded those of CPUs released at the
same time, with the recently announced Volta generation V100 Nvidia GPU promising a
single-precision floating point performance of up to 15 TFLOPs per device. The main
reason for this lead in performance is a difference in design goals: current CPUs are
optimized to deliver the best possible serial performance under an unpredictable,
interactive load. To achieve this, a large proportion of the available die space is
devoted to pipelining, branch prediction, and similar control logic that helps to
improve single-thread performance, as well as a hierarchy of relatively large cache
memories that are required since locality of memory accesses under a mixed
interactive load cannot be ensured. GPU dies of the same complexity, on the other
hand, feature a much larger number of actual compute units, much lighter control
logic and smaller cache memories. In an interactive load situation they would not
perform well, but for repetitive and highly parallel calculations they can deliver
exceptional performance. This makes them ideal vehicles for general-purpose
scientific calculations (GPGPU) \cite{owens:08}.

\begin{figure}[tb]
  \centering
  \includegraphics[clip=true,keepaspectratio=true,width=0.8\columnwidth]{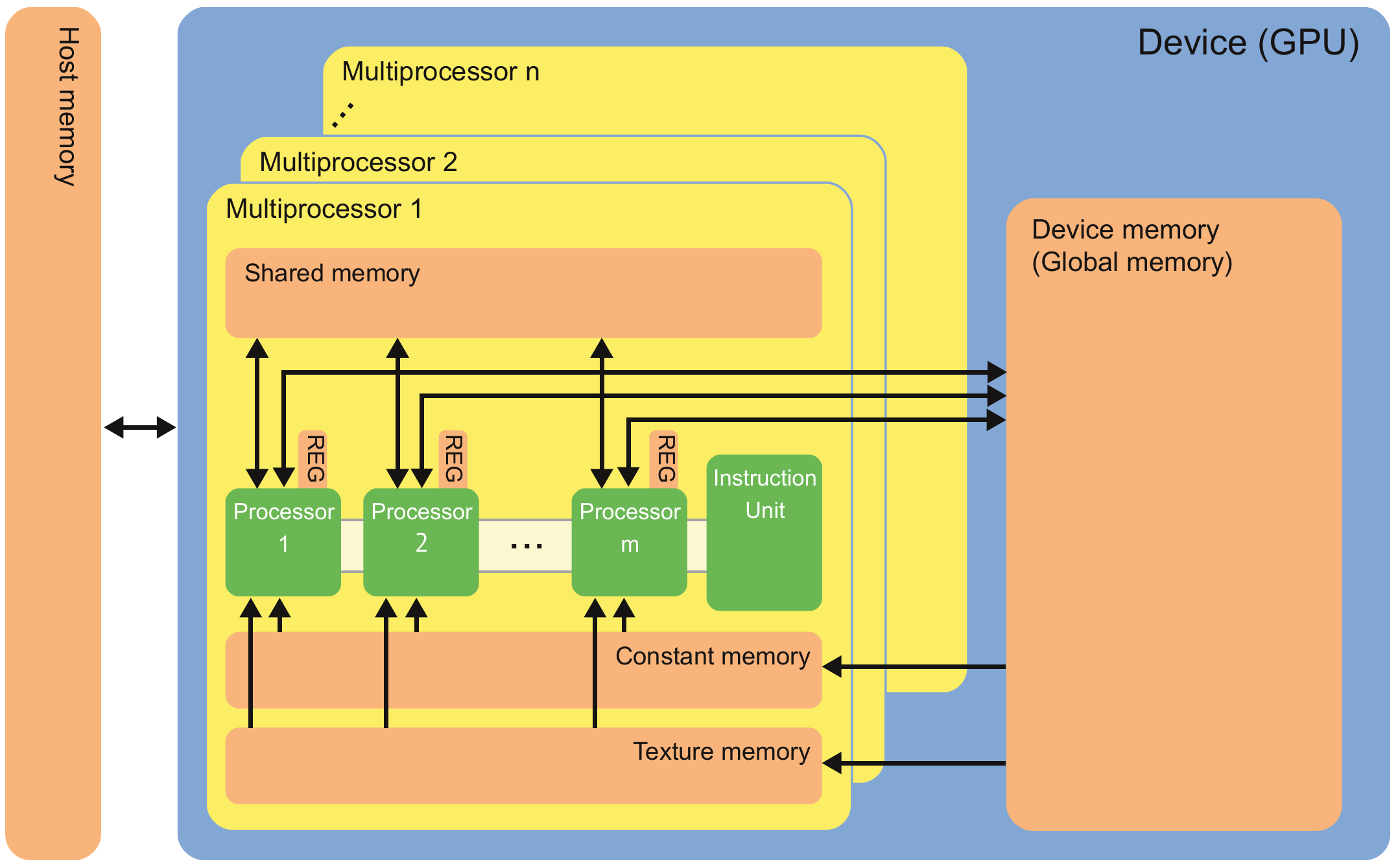}
  \caption
  {Schematic of the architecture of GPU devices, using the terminology of
    Nvidia CUDA.}
  \label{fig:hardware}
\end{figure}

The two main players in the high-end GPU market are Nvidia and AMD. Although the
actual performances of corresponding boards from both vendors are about similar,
Nvidia GPUs are much more firmly established as accelerator devices in HPC. This is,
in part, due to a rather well developed eco-system of development tools, and
supporting application libraries. The standard model for programming Nvidia GPUs is
through the proprietary CUDA framework \cite{cuda}, providing a C/C++ language
extension with the associated compiler, performance analysis tools and application
libraries. Less machine specific frameworks such as \index{OpenCL} OpenCL and OpenACC
are also available, and can also be used for programming AMD GPUs. Due to limitations
in the accessible features and a lack of fine-grained control they are somewhat less
popular for Nvidia GPUs applied in HPC, but they provide portable code that can run
on GPUs of different vendors and even multi-core CPU systems.  Figure
\ref{fig:hardware} shows a schematic of the general layout of a GPU device. It
consists of a number of multi-core processors (known as ``streaming multiprocessors''
for Nvidia devices) with the associated local, {\em shared memory\/}
\index{memory!shared} and a common {\em global memory\/} \index{memory!global} per
device. The number of cores per multiprocessor is between 32 and 192 in Nvidia cards
ranging through the Fermi, Kepler, Maxwell, and Pascal series, and each GPU card
comes with a few tens of multiprocessors, thus totaling in several thousand cores for
the larger cards. The associated compute model features elements of \index{SIMD} SIMD
and MIMD \index{MIMD} systems which sometimes is called single instruction, multiple
threads \index{SIMT} (SIMT). It corresponds to a tiled SIMD architecture, where each
multiprocessor has SIMD semantics, but the vector lanes are promoted to fibers with
the possibility of divergent control flow through the masking out of lanes for
branches that they do not take. These threads are very lightweight and the overhead
for their scheduling is minimal. As groups of 32 threads (a {\em warp\/})
\index{CUDA!warp} are scheduled together on a single SIMD processor, it is important
to minimize \index{thread~divergence} thread divergence using masking as it severely
impedes performance \cite{kirk:10}.

\begin{figure}[tb]
  \centering
  \includegraphics[clip=true,keepaspectratio=true,width=0.8\columnwidth]{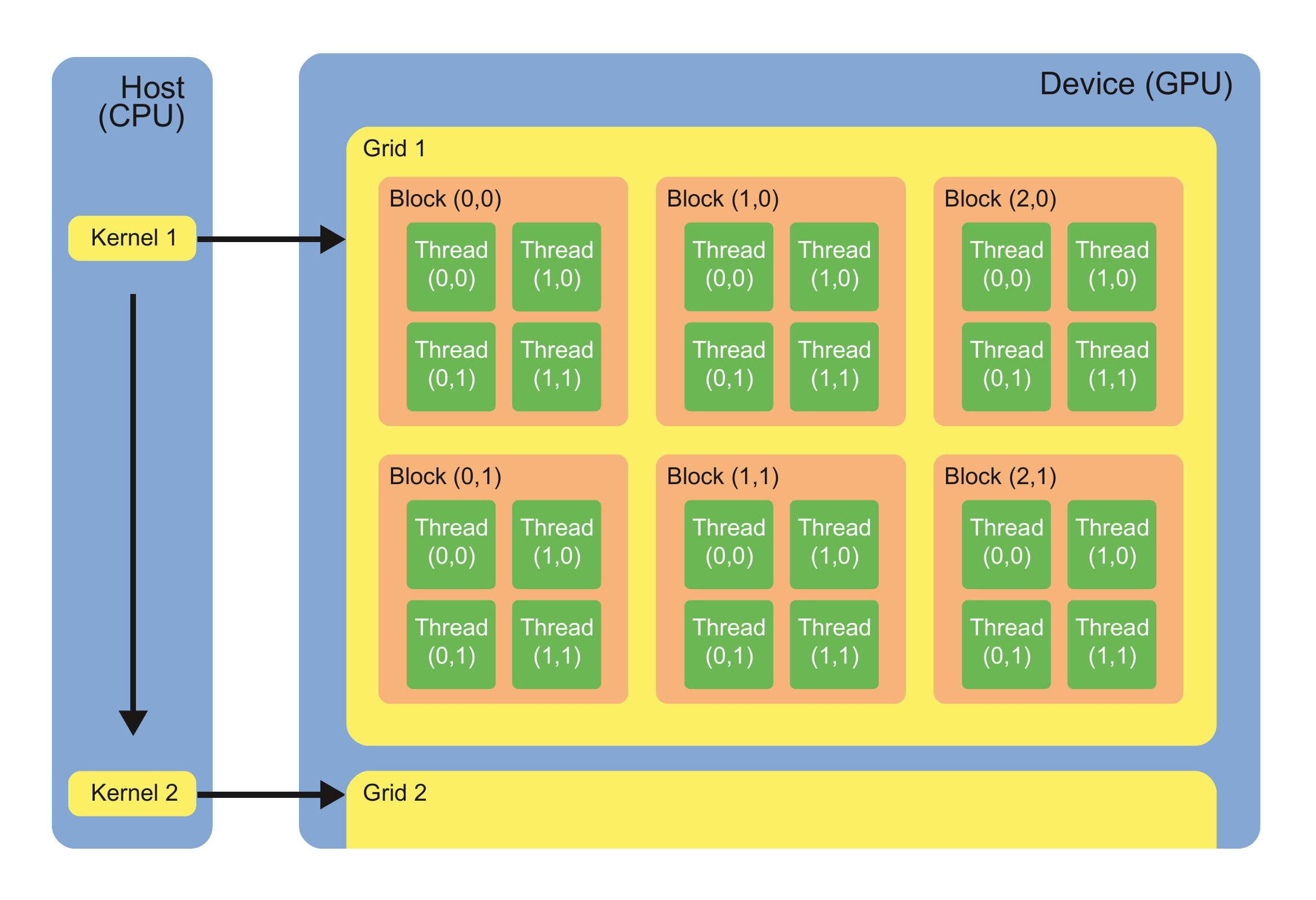}
  \caption
  {Execution configuration of device code in the form of a grid of thread blocks.}
  \label{fig:gridblock}
\end{figure}

\index{CUDA|(}

The main control flow of programs in CUDA (and similarly for OpenCL) is executed on
CPU, and it contains particularly labeled device functions \index{CUDA!kernel} (known
as {\em kernels\/}) that offload specific calculations to the GPU device. The
particular arrangement of threads to be used for a kernel invocation is known as the
\index{CUDA!execution~configuration} {\em execution configuration\/}, and it
describes a grid \index{CUDA!grid} of \index{CUDA!thread~block} thread blocks. This
is illustrated in Fig.~\ref{fig:gridblock}. Each block is scheduled to execute on a
single SIMD processor. Its threads can communicate (and synchronize) via the
\index{memory!shared} shared memory area local to it. Threads in different blocks
cannot directly communicate, and synchronization of all threads in a grid can only
happen through returning to CPU \index{atomic~operation} code.\footnote{Note that
  there are some advanced features allowing for limited communication between
  different blocks from within a kernel, including atomic operations as well as
  memory fence functions. For details see the CUDA C Programming Guide
  \cite{cuda-guide}.} More recent cards and driver versions also allow for dynamic
parallelism, where additional threads can be spawned from within kernel code
\cite{cuda-guide}. The most important available memories are illustrated in
Fig.~\ref{fig:hardware}. Moving from registers through shared memory to global
memory, the latencies for accesses increase dramatically and the bandwidths decrease
correspondingly. Additionally, Nvidia cards feature L1 and L2 caches also. While
these have the usual associative cache line behavior, shared memory is allocated and
managed explicitly by the threads in a block. It is fast but very small, at most
48~KB per block, so must be used wisely. GPU devices can only be run as accelerators
attached to a CPU node. They are connected through the PCI-e bus and any data that is
required as input or output must be transferred from the CPU to the GPU main memory
explicitly. This is of particular importance for hybrid codes that perform part of
the calculations on CPU and for multi-GPU programs. Such transfers can be interleaved
with calculations, however, and in some cases this allows to completely hide these
memory transfer latencies. As an extension, it is also possible to enable a unified
virtual address space, such that the same pointers can be used across CPU and GPU
memories, and data are automatically transferred across the PCI bus as required by
the access patterns.

\index{CUDA|)}

It is not possible in the framework of the present chapter to provide a comprehensive
discussion of programming models for GPUs and, in particular, give an introduction to
the CUDA or OpenCL language extensions. A number of good books and online resources
fill this gap, see, e.g., Refs.~[\refcite{cuda-guide,cuda-opt,kirk:10,scarpino:12}]. It
will suffice for the present purposes to provide a list of issues to consider in
order to achieve good performance, roughly in the order of their relevance:
\begin{itemize}
\index{memory!coalescence|(}%
\item {\bf Memory coalescence}: each cached access to global memory fetches or writes
  a full cache line of 128 bytes. If only a single 4-byte word of this data is
  actually used, the efficiency of bus usage is extremely poor. In ideal access
  patterns, the 32 threads of a warp access memory locations in the same 128 byte
  cache line, leading to 100~\% efficiency of memory accesses. The actual performance
  penalty for misalignment depends on whether accesses are cached in L1 (128 byte
  cache lines) or in L2 only (32 byte segments). Improving memory coalescence implies
  good data locality \index{memory!locality} and will typically involve rearranging
  multi-dimensional arrays in memory such that the fastest-changing index corresponds
  to neighboring threads in a block.%
  \index{memory!coalescence|)} \index{parallel~slack|(}
\item {\bf Parallel slack}: accesses to shared memory typically take tens of clock
  cycles, accesses to global memory at least hundreds of cycles. These memory
  latencies can be hidden away by the scheduler if there is enough parallel
  slack. Once a warp issues a high-latency memory transaction, it is taken off the
  compute units into a dormant state and another warp with completed memory
  transaction is activated instead. As a rule of thumb, optimal performance is
  achieved with a parallel slack of at least 8--10 times the number of available
  cores, implying several ten thousand threads per grid on high-end GPUs.%
\index{parallel~slack|)}
\index{occupancy|(}
\item {\bf Occupancy}: There are limits to the total number of resident threads (2048
  on recent Nvidia cards) and the total number of blocks (16--32) per SIMD
  processor. Additionally, the number of registers per thread requested by a given
  kernel can further limit the total number of threads and blocks that can be
  assigned to a multiprocessor at any given time. To the extent that a limitation in
  register usage does not impede performance, maximum throughput is typically
  achieved by maximizing the occupancy of threads on each SIMD processor.\footnote{The
    CUDA toolkit provides an occupancy calculator spreadsheet to help determine the
    right parameters \cite{cuda}.} Also, it is generally best to choose the total
  number of blocks to be a multiple of the number of SIMD processors of the device
  used.
\index{occupancy|)}
\index{memory!shared|(}
\item {\bf Shared memory}: explicit caching of data in shared memory can lead to
  massive performance improvements compared to direct accesses to global
  memory. Advantages will be larger the more often data loaded into shared memory are
  reused. A common pattern is to load a tile of the system into shared memory, update
  it there and then write the result back to global memory. For best performance, the
  different threads in a warp need to access shared memory locations in different
  banks to avoid bank conflicts \cite{cuda-opt}.
\index{memory!shared|)}
\index{arithmetic~density|(}
\item {\bf Arithmetic density and data compression}: as the arithmetic peak
  performance of GPU devices is enormous, many codes are limited by the practically
  achieved bandwidth of memory transfers, i.e., moving the data to and from the
  compute units. If this is the case, the optimization strategy must be a combination
  of improving memory throughput and reducing the amount of data that needs to be
  transferred. Throughput can be mainly improved by ensuring coalescence of memory
  accesses. A reduction of memory transfers results from a good use of shared memory
  and caches, but also from the most compact storage of data. If, for instance, a
  dynamic degree of freedom is an Ising spin corresponding to one bit of information,
  it is wasteful on memory bandwidth to store it in a 32-bit word. Instead, it should
  be stored in an 8-bit variable or several spins should be packed as individual bits
  into a longer word. In a situation where performance is memory bound, it should be
  attempted to increase the arithmetic intensity of the relevant kernel. For
  instance, it can be beneficial to recalculate intermediate results instead of
  re-reading them from (or even storing them in) main memory.
\index{arithmetic~density|)}
\index{floating-point~performance|(}
\item {\bf Floating-point calculations}: Floating-point operations in double
  precision are significantly more expensive on GPUs than single-precision
  calculations. The typical performance penalty for using double precision ranges
  between two and eight on recent Nvidia cards, where the best double precision
  performance is only available on the much more expensive Tesla series of GPGPU
  cards, but not on the otherwise very similar gaming and consumer cards.  On CPUs
  such effects are typically not as pronounced. In practice the speed of many
  programs will not be only determined by floating-point performance, such that the
  overhead for using double precision might be less dramatic than indicated above. In
  general sticking to single precision or some form of mixed precision calculations,
  where some intermediate results are stored in single (or even half) precision and
  only sums over large numbers of elements use double (or higher) precision, can be
  useful strategies. Another aspect of floating-point performance on GPUs is the
  availability of hardware units for the evaluation of certain special functions such
  as square roots, exponentials, logarithms and trigonometric functions in single
  precision \cite{kirk:10}. These have somewhat reduced precision but much higher
  performance than the software versions \cite{weigel:10a}.
\index{floating-point~performance|)}
\index{thread~divergence|(}
\item {\bf Thread divergence}: The individual SIMD processors emulate threads by
  masking out vector lanes to which a certain code branch does not apply. For a
  conditional, this means that all branches are evaluated serially with all threads
  to which the current branch does not apply masked out. Having $n$ branches with the
  same computational effort hence increases the worst-case total runtime (at least)
  by a factor of $n$. Since scheduling happens on the level of \index{CUDA!warp}
  warps (of 32 threads on Nvidia cards), it will improve performance if it can be
  ensured that all threads of a warp take the same execution branch as otherwise the
  different paths will be serialized. In particular, one should avoid the use of
  block-wide thread synchronization in divergent code as it will slow down the
  execution of all warps.
\end{itemize}
\index{thread~divergence|)}
As we shall see for some of the examples below, taking the above optimization
considerations into account can turn a very moderate GPU speedup against serial code
which is comparable to that achievable by parallelizing the CPU code into a several
hundredfold speedup against the serial program.

\index{GPU|)}
\index{algorithmic~pattern|(}

\subsection{Algorithmic patterns}
\label{sec:patterns}

It is not possible within the scope of the present chapter to discuss in detail
general parallel algorithms and their implementation with the help of the available
language extensions such as MPI, Cilk Plus, or CUDA. To help avoid running into the
ubiquitous {\em serial traps\/}, i.e., unnecessary assumptions in coding deriving
from the general serial execution assumption commonplace until recently, and to ease
the transition from a serial to a parallel mindset required now for practitioners
developing computer simulation codes, it appears useful, however, to provide an
overview of the most pertinent general algorithmic {\em patterns\/} or algorithm
skeletons \cite{aldinucci:07}. To this end we follow closely the excellent exposition
in Ref.~[\refcite{mccool:12}].

The most basic pattern, which applies to serial and parallel programs alike, is the
ability to stack patterns, i.e., to replace a certain task in an algorithmic pattern
by another pattern and to do so hierarchically up to an essentially arbitrary
recursion depth. This ability, which is equivalent to the composability of functions
in \index{composability} mathematics, is called \index{nesting|see{composability}}
{\em nesting\/}. In serial computing, nesting is mostly straightforward: the body of
a loop, for example, can contain another loop or a conditional statement etc. In
parallel algorithms problems can arise when the nesting is allowed to be dynamic,
i.e., it grows with the size of the problem. This can create an unbounded number of
parallel threads as the input size increases, such that efficient implementations
need to decouple the \index{parallelism!potential} potential parallelism resulting
from the nested algorithm from the actually available hardware parallelism.

\index{algorithmic~pattern!control~flow|(}

\subsubsection{Control flow}

A natural distinction arises between control flow patterns and data management
patterns. Regarding control flow, serial patterns are rather straightforward and
mostly correspond to the elementary features available in most (procedural)
languages. The \index{algorithmic~pattern!sequence} {\em sequence\/} pattern
expresses the sequential execution of several tasks. Although there might not
actually be any dependence between the elements of a sequence, a serial program will
always execute them in the given order. Optimization phases of compilers and even the
control logic of modern processors (``out-of-order-execution'') in some cases will
change the order of execution in a sequence, however, if their analysis allows to
ascertain that the results will be unchanged. The
\index{algorithmic~pattern!selection} {\em selection\/} pattern corresponds to
conditional execution, usually expressed in an if statement. {\em Iteration\/}
\index{algorithmic~pattern!iteration} is the main serial pattern for accommodating
variably-sized inputs. A common strategy for parallel computing is the (possibly
automatic) parallelization of loops, but in many cases the simplest approaches fail
due to data dependencies \index{dependency} between the iterations. Finally, the
\index{algorithmic~pattern!recursion} {\em recursion\/} pattern (which is absent in
some languages such as Fortran 77), can often (but not always) be expressed by
iteration also, but sometimes allows for much simpler code. It is the natural match
for divide-and-conquer strategies.

Parallel control flow patterns are not quite as universally well known. The basic
examples generalize the serial patterns discussed above. If a sequence of tasks is
actually independent, {\em fork-join\/} \index{algorithmic~pattern!fork-join} can be
used to run them in parallel. After their completion execution returns to a single
thread. Typically, such independence is only relative, however, and the results of a
task are needed at some point later on in the program, such that some communication
of forked threads is required. Typically this is through synchronization
\index{synchronization} points (barriers), where all threads of a certain fork point
need to have completed a certain part (or all) of their task. Other parallel control
patterns are mostly generalizations of the iteration mechanism. The most important is
\index{algorithmic~pattern!map} {\em map\/}, where a function is applied to each
element of an index set. This corresponds to a serial loop where each iteration is
independent of the others, which is the case that is also handled well by
compiler-level parallelization. If the strict independence of elemental operations is
relaxed, and each application of the function has access also to certain neighboring
elements in the input vector, one speaks of a \index{algorithmic~pattern!stencil}
{\em stencil\/}. Here, decomposition of the vector into independent sub-sets (such as
for the checkerboard decomposition \index{domain~decomposition!checkerboard}
discussed below in Sec.~\ref{sec:checkerboard}) and tiling are important optimization
strategies. The stencil pattern is the work horse of most simulation codes, ranging
from lattice systems to computational fluid dynamics. In a
\index{algorithmic~pattern!reduction} {\em reduction\/}, the results of applying a
function to each element in a set are combined together. The most common combiner
functions used here are addition, multiplication, and maximum. Whether the combiner
function is associative and/or commutative decides to which degree the result depends
on the actual schedule of parallel operations. A typical parallel implementation
leads to a tree structure, where partial reductions are formed at each level and
passed down to the next level for further reduction. A combination of map and
reduction is given by the {\em scan\/} \index{algorithmic~pattern!scan} operation,
where for each position in the output a partial reduction of the input up to that
point is needed. To parallelize it, often additional intermediate calculations are
required, thus increasing the total work and possibly limiting the scaling properties
of the whole code. Finally, a {\em recurrence\/}
\index{algorithmic~pattern!recurrence} is a generalization of map and stencil where
each iteration can read {\em and write\/} neighboring elements in the vector.

\index{algorithmic~pattern!control~flow|)}
\index{algorithmic~pattern!data~management|(}

\subsubsection{Data management}

The allocation mechanism used for automatic variables and also for local variables in
function calls is \index{memory!stack~allocation} {\em stack allocation\/}. Since it
follows strict last in, first out (LIFO) logic, allocation and deallocation are
achieved simply through the stepping of a pointer and all stack data is contiguous in
memory. For dynamic or \index{memory!heap~allocation} {\em heap allocation\/}, on the
other hand, there is no prescribed order of allocation and deallocation operations
and, as a consequence, the locations of consecutive allocations can be become
scattered over distant parts of the actual physical memory, thus limiting performance
where it depends on memory locality. In languages that allow it, memory accesses are
often through direct read and write accesses using pointers. These can make (in
particular automatic) vectorization and parallelization difficult as it typically
cannot be ascertained at compile time whether two different pointers refer to the
same location in memory or not (a problem known as aliasing).

In parallel codes, data locality \index{memory!locality} is particularly
important. In a distributed memory setup, clearly accesses to local memory will be
substantially more efficient than requesting data from a different MPI \index{MPI}
node. An even more fundamental problem is the concurrency of accesses, in particular
to avoid race conditions \index{memory!race~condition} resulting from uncontrolled
interleaved read and write accesses to the same locations. In general, it is
important to understand whether a given data element is shared between different
workers and when it is not and, as a result, to place it into a memory with the
appropriate scope. Some parallel data patterns include
\index{algorithmic~pattern!pack} {\em pack\/}, where a subset of an input vector
selected by another, Boolean selection vector of zeros and ones is placed next to
each other in a contiguous fashion; \index{algorithmic~pattern!pipeline} {\em
  pipelines\/}, where different stages in a sequence of operations run independently
as separate threads each of which delivers partially processed data to the next
stage; the \index{algorithmic~pattern!geometric~decomposition} {\em geometric
  decomposition\/} mentioned above in the context of the stencil pattern that uses
tiles, strips, checkerboards or other suitable geometric domains to be worked on in
parallel; and the \index{algorithmic~pattern!scatter} {\em gather\/} and
\index{algorithmic~pattern!gather} {\em scatter\/} pair of operations that use a data
vector and a set of indices and either reads (gather) or writes (scatter) in the data
vector at the locations given in the index vector.

More advanced patterns such as superscalar \index{parallelism!superscalar} sequences
or branch-and-bound are beyond the scope of the present introduction. They are
described in detail in Refs.~[\refcite{mattson:04,mccool:12}].

\index{algorithmic~pattern!data~management|)}
\index{algorithmic~pattern|)}

\subsubsection{Pitfalls}

Before discussing the actual applications of massively parallel computing in computer
simulations in statistical physics, it is perhaps useful to summarize again the most
common pitfalls of parallel algorithms and the basic approaches for avoiding them.

\index{memory!race~condition|(}

{\em Race conditions\/} are among the most common and difficult to debug problems in
parallel codes. If, for example, two threads try to increment a shared variable, one
of the updates can be lost if the read of the second thread occurs after the read of
the first thread but before the write operation of the first thread. As the results
depend on the typically unpredictable order of execution of individual threads, these
problems are often intermittent in nature. If at all possible, the potential for such
races should be avoided by choosing suitable algorithms. If this is not possible,
races can be tamed by the use of memory fences and locks, which essentially guarantee
one set of operations to be finished before the other set starts. Operations on locks
must occur atomically, such that they appear instantaneous to other tasks.

\index{memory!race~condition|)}
\index{deadlock|(}

{\em Deadlock\/} occurs when two or more tasks are waiting for each other to complete
certain tasks and each cannot resume until the other task finishes. This can be the
case, for example, if several locks need to be acquired by more than one task, each
task acquires one of the locks and waits for the other one to become available. It
can be avoided if locks are always acquired by all tasks in the same order, but the
problem serves to show that locks are best avoided. Locks also create serial
bottlenecks in the code as all operations on a single lock must occur in sequential
order. This effect will impede the scaling of an algorithm, but whether this is
practically relevant depends on the frequency of use of the lock and the actual
number of threads employed.

\index{deadlock|)}
\index{memory!locality|(}

The other main pitfall in parallel code is a {\em lack of locality\/}. Most hardware
is built on the assumption that for each memory transaction each core is likely to
either use the same or a nearby memory location again in the nearby future (temporal
and spatial locality). To avoid problems in this respect, parallel code needs to use
a suitable layout of data in memory and make good use of cache memories where they
are available.  On GPUs this includes the issues of coalescence of memory
transactions, the use of shared memory \index{memory!shared} and an appropriate cache
configuration as discussed above in Sec.~\ref{sec:hardware}.

\index{memory!locality|)}
\index{load~imbalance|(}

Depending on the parallelization strategy and the nature of the problem, another
source of inefficiency arises from {\em load imbalance\/} between parallel
threads. Apart from suitably changing the parallelization strategy, a fine-grained
decomposition of work can help to mitigate the effects of load imbalance. Also,
adaptive schemes of idle threads acquiring new work via a scheduler can lead to
improvements here. If, on the other hand, the over-decomposition of work is pushed
too far, there is a danger of the parallel {\em overhead\/} for thread
initialization, copying of data etc. to outweigh the scaling gain, especially if the
arithmetic intensity of individual threads becomes too low.

\index{load~imbalance|)}

\section{Canonical Monte Carlo}
\label{sec:canonical}

\index{Markov~chain|(}

There is by now a very wide range of Monte Carlo methods that are used for
simulations of systems in (classical) statistical physics
\cite{binder:book2,frenkel:02}. While there are a few exceptions (and we will discuss
one below in Sec.~\ref{sec:PA}), the overwhelming majority of methods are based on
Markov chain Monte Carlo (MCMC) that allows to implement importance sampling and also
simulations in generalized ensembles \cite{berg:04}. In this scheme, configurations
are modified in each step according to transition probabilities that only depend on
the current configuration. The resulting Markov chain of configurations,
\[
  \{s_i\} \to \{s_i'\} \to \{s_i''\} \to \ldots
\]
where the $s_i$ denote the configurational variables, converges to a stationary
distribution $\pi(\{s_i\})$ if the chosen move set is \index{ergodicity} {\em
  ergodic\/} (i.e., loosely speaking, it allows to connect all pairs of states within
a finite number of steps) and the transition probabilities $T(\{s_i\} \to \{s_i'\})$
satisfy the {\em balance\/} equation
\begin{equation}
  \label{eq:balance}
  \sum_{\{s_i'\}} \pi(\{s_i\}) T(\{s_i\} \to \{s_i'\}) =
  \sum_{\{s_i'\}} \pi(\{s_i'\}) T(\{s_i'\} \to \{s_i\}).
\end{equation}
The simplest way of fulfilling Eq.~(\ref{eq:balance}) is to demand equality term by
term under the sums,
\begin{equation}
  \label{eq:DB}
    \pi(\{s_i\}) T(\{s_i\} \to \{s_i'\}) = \pi(\{s_i'\}) T(\{s_i'\} \to \{s_i\}).
\end{equation}
This is known as {\em detailed balance\/} \index{detailed~balance}
condition. Together with ergodicity it is sufficient, but in contrast to balance,
Eq.~(\ref{eq:balance}), it is not necessary to ensure convergence.

There is some further freedom in implementing Eq.~(\ref{eq:DB}). The best known
approach is the \index{Metropolis~algorithm} Metropolis algorithm
\cite{metropolis:53a} where
\begin{equation}
  \label{eq:metropolis}
  T(\{s_i\} \to \{s_i'\}) = \min\left[1,\frac{\pi(\{s_i'\})}{\pi(\{s_i\})}\right].
\end{equation}
In the simplest method, the proposed configuration $\{s_i\}'$ only differs from the
current one $\{s_i\}$ in a single degree of freedom, for example the orientation of a
spin or the position of a particle. While the scheme is also valid for any other
modification rule, any sufficiently non-local update --- unless ingeniously crafted
\cite{swendsen-wang:87a} --- will result in largely different probabilities
$\pi(\{s_i\})$ and $\pi(\{s_i'\})$ and hence very small move acceptance rates. For
the actual implementation it is often useful to decompose the transition probability
as
$T(\{s_i\} \to \{s_i'\}) = C(\{s_i'\} | \{s_i\})\,p_\mathrm{acc} (\{s_i'\} |
\{s_i\})$, where $C$ is the proposal probability for a certain move
$\{s_i\} \to \{s_i'\}$ and $p_\mathrm{acc}$ is a move acceptance probability
evaluated according to Eq.~(\ref{eq:metropolis}). The proposal probability $C$
determines the order in which individual degrees of freedom are tried, the most
common approaches being, respectively, a uniformly random spin selection and a
sequential selection of spins in successive steps, traversing the lattice in a
regular fashion.

\index{Markov~chain|)}
\index{heatbath~algorithm|(}

Another standard approach for satisfying the detailed balance condition (\ref{eq:DB})
is the heatbath method for the update of a single variable $s_k$, where its new value
is directly chosen from the equilibrium distribution $\pi$ conditioned on the given
values of the remaining degrees of freedom $s_j$, $j\ne k$:
\begin{equation}
  \label{eq:heatbath}
  T(\{s_i\} \to \{s_i'\}) =
  \frac{\pi(s_k' | \{s_{j\ne k}\})}{\sum_{s_k} \pi(s_k' | \{s_{j\ne k}\})}.
\end{equation}
If $s_k'$ only takes values from a finite set of options, sampling from the above
distribution is straightforward by using geometric sampling from the cumulative
distribution \cite{janke:08} $\sum_{s_k'=s_\mathrm{min}}^{s_\mathrm{max}} \pi$ or by
using more advanced techniques such as Walker's method of alias
\cite{loison:04,fukui:09}. For continuous degrees of freedom the method can be
implemented if there is an analytical inversion of the cumulative distribution
function corresponding to Eq.~(\ref{eq:heatbath}) \cite{miyatake:86} or by using
tables to approximate this expression \cite{loison:04}.

\index{heatbath~algorithm|)}

For the examples in this section we will focus on simulations in the canonical
ensemble with
\[
\pi(\{s_i\}) = \frac{1}{Z_\beta} \exp[-\beta{\cal H}(\{s_i\})],
\]
where ${\cal H}(\{s_i\})$ is the Hamiltonian and $\beta$ denotes inverse
temperature. Note that the partition function $Z_\beta$, that is in general unknown,
drops out of the expressions (\ref{eq:metropolis}) and (\ref{eq:heatbath}) for the
transition probabilities. Other ensembles, such as $NpT$ or $\mu V T$ for particle
systems, can be realized in a similar way, and we discuss generalized-ensemble
simulations in Sec.~\ref{sec:generalized} below.

\index{Ising~model|(}

For definiteness, we first focus on the nearest-neighbor Ising model with Hamiltonian
\begin{equation}
  \label{eq:ising_model}
  {\cal H} = -\sum_{\langle i,j\rangle} J_{ij} s_i s_j - \sum_i h_i s_i.
\end{equation}
Here, $J_{ij}$ are the exchange couplings between nearest-neighbor spins and $h_i$
denotes an external magnetic field acting on the spin $s_i$. We initially concentrate
on the ferromagnetic model with uniform couplings $J_{ij} = J = 1$ and in the absence
of magnetic fields, $h_i = 0$, and come back to the case of disordered systems in
Sec.~\ref{sec:disordered}. For the purposes of the present chapter, we will always
apply periodic boundary conditions as is typically recommended to minimize
finite-size effects, but other boundary conditions can be implemented quite easily as
well, and the optimizations mentioned here are essentially independent of this
choice.  Generalizations to models with different finite interaction ranges are
rather straightforward and only lead to different decompositions of the lattices into
interpenetrating sub-lattices of non-interacting sites. Systems with truly long-range
interactions require different methods which are outside of the scope of the present
discussion \cite{flores:17,michel:17}.

\index{Ising~model|)}

\subsection{Checkerboard scheme}
\label{sec:checkerboard}

Following the algorithm outlined above, a single spin-flip simulation of the Ising model
(\ref{eq:ising_model}) with the Metropolis algorithm comprises the following steps:
\begin{enumerate}
\item Initialize the system, possibly with a uniformly random spin configuration.
\item Choose the lattice site $k$ to update, according to the scheme used, either randomly,
  sequentially or in a checkerboard fashion.
\item Calculate the energy change incurred by flipping spin $k$,
  \[
    \Delta E_k = 2 s_k \sum_{j\,\mathrm{nn}\,k} s_j.
  \]
  Draw a random number $r$ uniformly in $[0,1)$. Accept the flip if $\Delta E_k \le
  0$ or 
  \begin{equation}
    r < \exp(-\beta\Delta E_k),
    \label{eq:metropolis-canonical}
  \end{equation}
  otherwise reject and maintain the current configuration as the new state.
\item Repeat from step 2 until the prescribed number of updates has been completed.
\end{enumerate}
In a serial implementation, typically random or sequential site selection would be
used. Sequential updates lead to somewhat faster relaxation of the chain
\cite{ren:06}, which can be understood qualitatively from the possibility to transmit
information about spin updates ballistically through the lattice in the direction of
sequential progression. Additionally, a sequential update is cheaper computationally
than visiting sites in random order as it features good memory locality and it also
does not require an additional random number for site selection. Note that sequential
updates do not satisfy detailed balance (while still satisfying balance)
\cite{berg:04}, which needs to be taken into account for studies of dynamical
properties.

\begin{figure}[tb]
  \centering
  \includegraphics[clip=true,keepaspectratio=true,width=0.4\columnwidth]{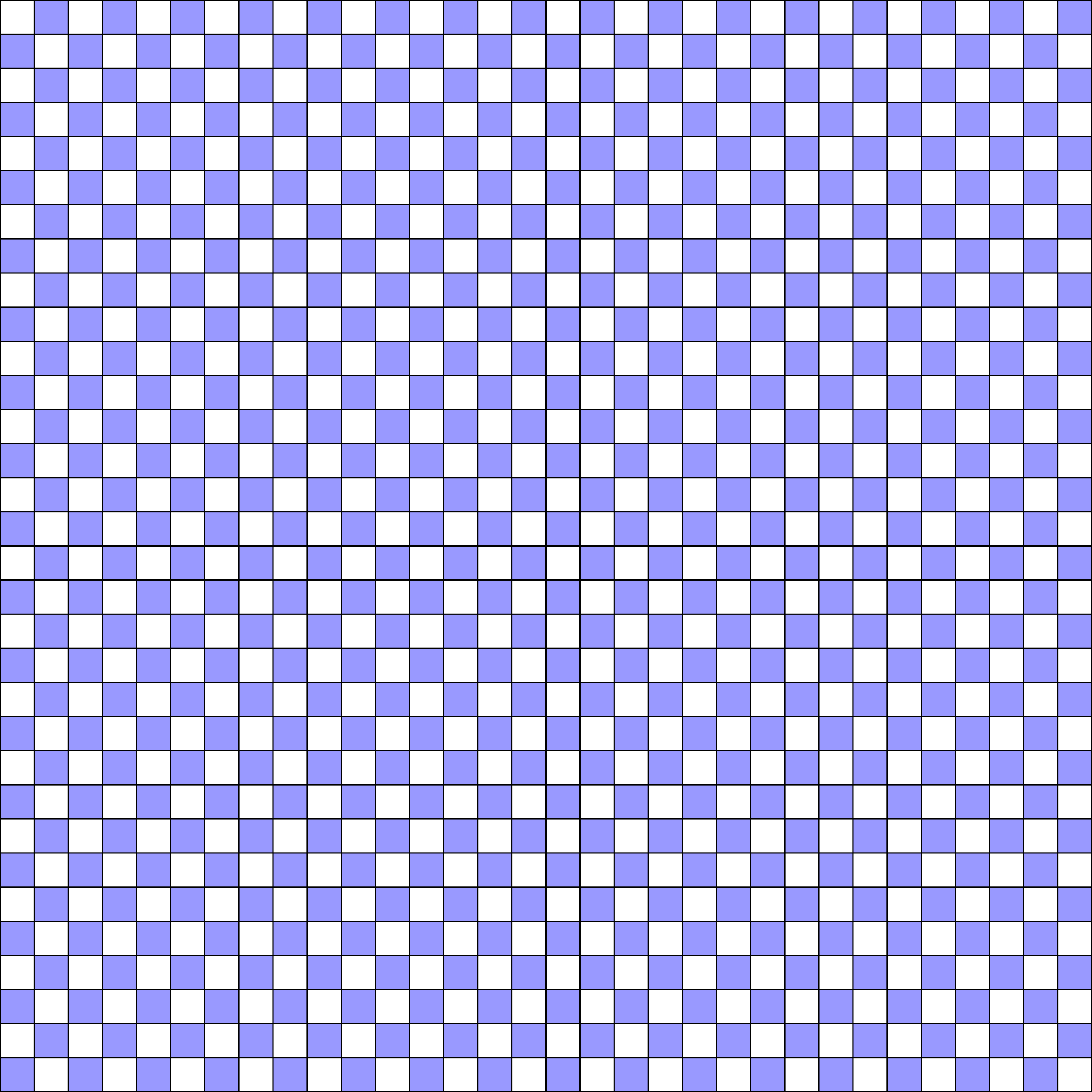}
  \hspace{0.5cm}
  \includegraphics[clip=true,keepaspectratio=true,width=0.4\columnwidth]{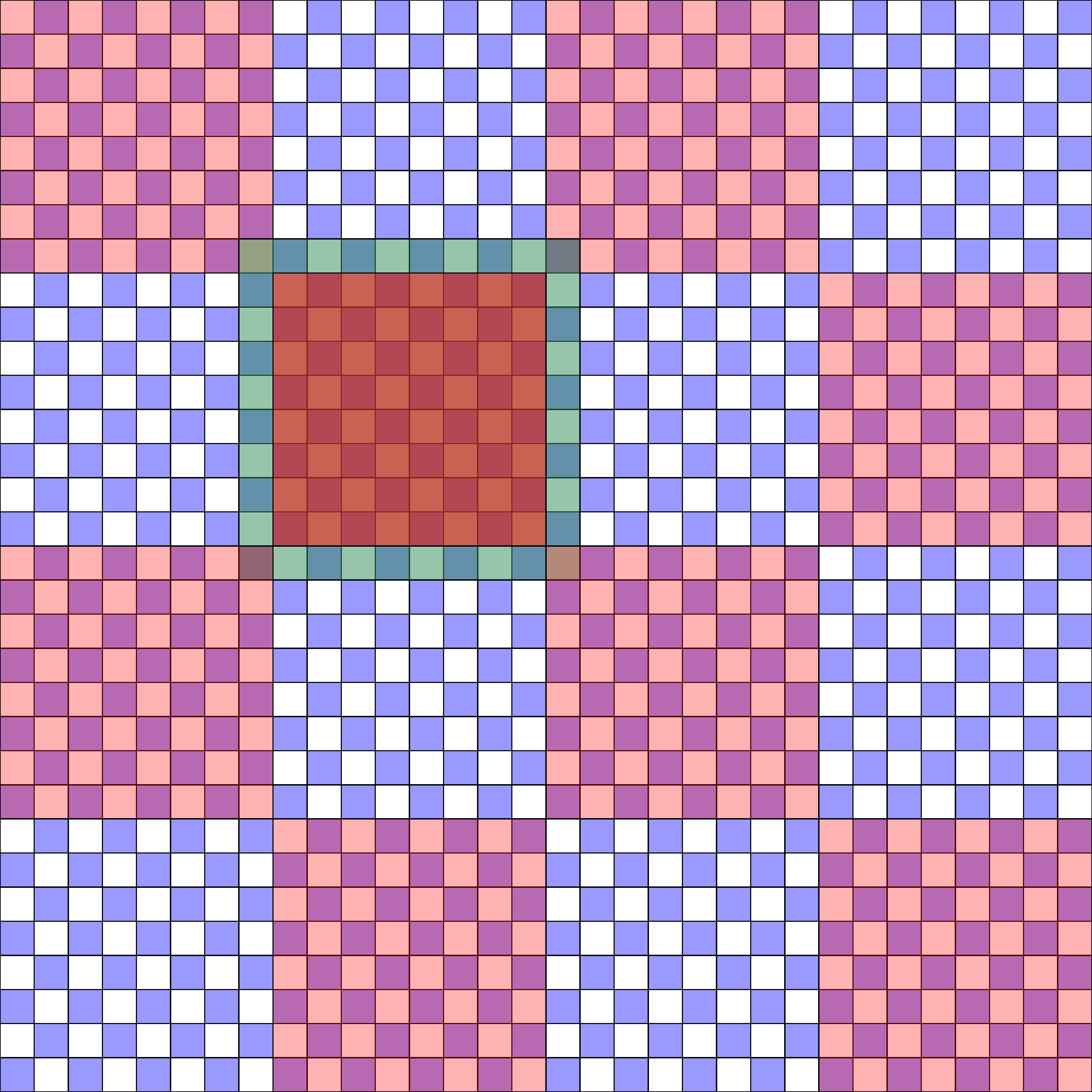}
  \caption
  {Left: checkerboard decomposition of the square lattice. Right: double checkerboard
    decomposition of the square lattice. The tiles of $8\times 8$ sites are assigned
    to thread blocks, and red (darker) and blue (lighter) tiles are updated in an
    alternating fashion. Within each tile, the threads of a block update spins of one
    sub-lattice, synchronize, and then update the other sub-lattice. The shaded tile
    and halo indicate the subset of spins that are cooperatively loaded into shared
    memory by the threads of a block.}
  \label{fig:checkerboard}
\end{figure}

\index{domain~decomposition!checkerboard|(}

For parallel updates, on the other hand, a suitable domain decomposition of the
lattice is required. For bipartite lattices and nearest-neighbor interactions this
results in a checkerboard (or generalized checkerboard in three and higher
dimensions) labeling of the lattice, which allows different spins on the same
sub-lattice to be updated in parallel, independent of each other. This is illustrated
for the square lattice in the left panel of Fig.~\ref{fig:checkerboard}. For
different lattice types and/or models with different (finite) interaction ranges,
similar decompositions can always be found, the only difference being that they in
general require more than two sub-lattices. A full sweep of spin updates in this
scheme then corresponds to a parallel update of all spins on the even sub-lattice
followed by a \index{synchronization} synchronization of all threads and a parallel
update of all spins of the odd sub-lattice. In the maximally parallelized version
each spin of one of the sub-lattices is updated by a separate thread, leading to a
total of $N/2$ parallel threads, where $N=L^d$ is the total number of spins. For a
GPU implementation, the restriction in the number of parallel threads in a block
(1024 for recent Nvidia GPUs) makes it necessary for all but the smallest systems to
decompose the lattice using tiles, for which one possibility in 2D is a square shape
with $T\times T$ spins each as shown in the right panel of
Fig.~\ref{fig:checkerboard}. Other shapes such as strips can also be used
\cite{preis:09}, and the optimal shape depends on the arrangement of spins in memory
and the caching mechanisms employed \cite{lulli:15}.

\index{domain~decomposition!checkerboard|)}

The resulting GPU simulation code is very simple as is apparent from the CUDA
implementation of the simulation kernel shown in Fig.~\ref{fig:metro-one}. The random
numbers required for implementing the Metropolis criterion are created via inline
instances of generators, one per thread, hidden behind the macro {\tt RAN(x)}. As an
evaluation of the exponential function in Eq.~(\ref{eq:metropolis-canonical}) is
relatively expensive computationally, it is common practice for systems with a small
number of states per spin to tabulate the possible values of
$\exp(-\beta \Delta E_k)$, and this was also done here with the result stored in the
array {\tt boltzD}. The sub-lattice is selected using the {\tt offset} variable that
should be either 0 or 1, such that the kernel needs to be called twice to achieve a
full update, once for each sub-lattice. In this case thread synchronization is
achieved through a return of the control to the CPU code in between kernel calls
(which are in the same stream), ensuring that all calculations of the first call have
completed before the second call is executed.

\begin{figure}[tb]
  \begin{lstlisting}
    __global__ void metro_checker(int *s, int *ranvec, int offset)
    {
      int y = blockIdx.y*BLOCKL+threadIdx.y;
      int x = blockIdx.x*BLOCKL+((threadIdx.y+offset)%2)+2*threadIdx.x;
      int xm = (x == 0) ? L-1 : x-1, xp = (x == L-1) ? 0 : x+1;
      int ym = (y == 0) ? L-1 : y-1, yp = (y == L-1) ? 0 : y+1;
      int n = (blockIdx.y*blockDim.y+threadIdx.y)*(L/2)+
               blockIdx.x*blockDim.x+threadIdx.x;
      
      int ide = s(x,y)*(s(xp,y)+s(xm,y)+s(x,yp)+s(x,ym));
      if(ide <= 0 || fabs(RAN(ranvec[n])*4.656612e-10f) < boltzD[ide])
        s(x,y) = -s(x,y);
    }
  \end{lstlisting}
  \caption{
    The simplest, still poorly optimized CUDA kernel for a GPU simulation of the 2D Ising
    model with the single spin-flip Metropolis update.
    \label{fig:metro-one}
  }
\end{figure}

The parallel speedup observed for this code run on a Tesla C1060 card with 240 cores
over a serial code run on a CPU of the same period (Intel Q9650) is about
10-fold. This rather moderate improvement is typical for many of the simplest
implementations that do not take many specifics of the architecture into account. In
view of the general performance guidelines sketched in Sec.~\ref{sec:hardware} a
number of improvements come to mind:
\begin{itemize}
  \index{memory!coalescence|(}
  \index{memory!locality|(}
\item The locality of memory accesses and hence coalescence is not very good in the
  setup of {\tt metro\_checker()}. Successive threads in a block update spins that
  are at least two memory locations apart (for spins in the same row) or even
  potentially arbitrarily far apart for spins in different rows of the same
  tile. Also, when calculating the sum over nearest neighbor spins in the variable
  {\tt ide} accesses are not coalesced and each spin is read twice as the right
  neighbor of a spin on the even sub-lattice (say) is the left neighbor of the next
  spin on the same sub-lattice etc. If the second reads are not served from a cache,
  they will be as expensive as the first ones. A natural improvement increasing the
  coalescence of operations on spins to be updated is to store the even and odd
  sub-lattices separate from each other, potentially arranging them tile by tile to
  avoid problems with non-locality of memory accesses in moving from row to row. A
  further improvement can be achieved by re-shuffling the spins in each sub-lattice
  in a way such that as many neighbors as possible of a spin on one sub-lattice
  appear as consecutive elements in the array for the other sub-lattice. A scheme
  dubbed ``crinkling'' that ensures that three out of four neighbors are next to each
  other for the square lattice was proposed in Ref.~[\refcite{ferrero:11}]. A
  similar, but more general scheme of ``slicing'' for hypercubic lattices in any
  dimension is used in Ref.~[\refcite{lulli:15}]. A ``shuffled'' \cite{fang:14} or
  ``interlaced'' \cite{manssen:15} memory layout combines the separate storage of odd
  and even sub-lattices for two realizations to further improve coalescence.  To
  reduce the arithmetic load incurred by the required index calculations for spin
  accesses, one might also bind the arrays for each sub-lattice to a
  \index{memory!texture} texture.\footnote{Textures are handled in a separate memory
    hierarchy equipped with additional hardware units for indexing \cite{cuda}.}
  \index{memory!coalescence|)} \index{memory!locality|)}
\item The Boltzmann factors are already tabulated in the array {\tt boltzD} to avoid
  the expensive evaluation of the exponential function. This part can be further sped
  up by using a texture for storing the array since textures are well optimized for
  read accesses of different threads of a warp to different locations (which will be
  the case since the energy changes $\Delta E_k$ will differ between spins).
\index{multi-spin~coding|(}
\item The use of {\tt int} variables of 32 or 64 bits is wasteful for the storage of
  the one-bit information $s_i$ and causes unnecessary data transfer over the bus
  that can slow down the code. It is straightforward to replace the {\tt int}s by
  only 8-bit wide {\tt char}s which already provides for a noticeable speedup. An
  additional improvement can be achieved by the use of {\em multi-spin coding\/}
  (MSC), typically implemented with {\tt int} variables of 32 or 64 bits, to ensure
  that each spin only occupies one bit of storage. For the simulation of a single
  ferromagnetic system, this means that spins located at different lattice sites need
  to be coded together, and it is typically most efficient to unpack them on GPU for
  the actual spin update. Some of the details are discussed in
  Ref.~[\refcite{block:10}]. To achieve results of high statistical quality, it is
  important in this setup to use independent random numbers for updating each of the
  spin coded in the same word. A related issue for a simulation method involving a
  population of configurations is discussed below in Sec.~\ref{sec:PA} in the context
  of the implementation of the population annealing algorithm.
\index{multi-spin~coding|)}
\item It is possible to explicitly disable the use of L1 cache for reads.\footnote{The
    relevant {\tt nvcc} compiler switch is {\tt -Xptxas -dlcm=cg}.} As a result a
  cache miss fetches a 32 byte segment and stores it in L2, whereas otherwise an L1
  cache line of 128 bytes would be loaded. For the ``crinkled'' memory setup this
  tends to increase memory bus efficiency for the neighbor that is not sequentially
  aligned.
  \index{thread~divergence|(}
\item On some cards it can be advantageous to remove thread divergence and ensure
  write coalescence by updating the spin variable irrespective of whether the flip
  was accepted and only deciding about the new orientation in a local variable in the
  Metropolis condition.
  \index{thread~divergence|)}
\end{itemize}
Storing the two sub-lattices separately, using the crinkling transformation, binding
the Boltzmann factors to a texture, using {\tt char}s to store spins, disabling L1
cache and using coalesced writes increases the speedup factor on the Tesla C10560 to
about 60. Further improvements can be achieved by using textures for the spin arrays
and further optimizations of the memory arrangements and access patterns as described
in detail in Ref.~[\refcite{lulli:15}]. We note that more recent cards are somewhat
less sensitive to data locality issues due to improved automatic caching. On the
Maxwell card GTX Titan Black, for example, we find spin-flip times of about 0.2 ns
for the initial version of the code, corresponding to an about 30-fold speedup
compared to an Intel Xeon E5-2620 v3 CPU, whereas the optimizations mentioned above
boost this performance to an about 100-fold speedup with $t_\mathrm{flip} = 0.06$~ns
($L=4096$).

\index{memory!shared|(}
An alternative optimization strategy lies in the use of shared memory by loading
tiles of the configurations into this fast cache:
\begin{itemize}
\item Storing the spin configuration of the tile that is currently being updated in
  shared memory allows to avoid problems with non-coalescence of global memory
  accesses as well as the double reads of neighboring spin orientations. To allow for
  parallel updates of the configuration in several tiles, these in turn need to be
  also arranged in a checkerboard fashion, leading to the two-level ``double
  checkerboard'' decomposition shown in the right panel of
  Fig.~\ref{fig:checkerboard} \cite{weigel:10c,weigel:10a}. In this setup, one
  requires $(L/T)^2/2$ thread blocks and each of them collectively updates one tile
  of the red shaded (coarse) sub-lattice shown in Fig.~\ref{fig:checkerboard}. After
  this kernel call has completed, a second call requests the same blocks to update
  the other coarse sub-lattice of tiles, thus completing a full sweep. Each block
  consists of $T^2/2$ threads that collectively load the configuration of the tile
  from global memory into shared memory, with some of the threads additionally
  loading the one spin wide halo around the tile required to update the boundary
  spins correctly. After the load, all threads of the block are synchronized and then
  update the spins in shared memory in a checkerboard fashion as in the version
  without shared memory. Note that the two sub-lattices of a single tile are now
  updated from within the same kernel call.
\item The full potential of this approach is only released if each tile, once loaded
  to shared memory, is subjected to several rounds of spin updates. If $k$ rounds of
  updates are performed, the resulting \index{multi-hit~updates} ``multi-hit'' code
  is particularly economic on memory transfers and hence is able to fully load the
  available computational units. This approach does not satisfy detailed balance, but
  the same applies to any checkerboard or sequential update, so this is no particular
  drawback of the method, but it needs to be taken into account when studying
  dynamical properties. It is clear, however, that very close to the critical point,
  the multi-hit approach will have slightly larger autocorrelation times than a
  single-hit variant as information can only be transmitted between tiles after each
  full update of the lattice. The resulting optimal choice of $k$ was studied in some
  detail in Ref.~[\refcite{weigel:11a}] and found to be around $k=10$ near
  criticality.
\end{itemize}
\index{memory!shared|)}

\begin{figure}[tb]
  \centering
  \includegraphics[keepaspectratio=true,scale=0.65,trim=45 48 75 78]{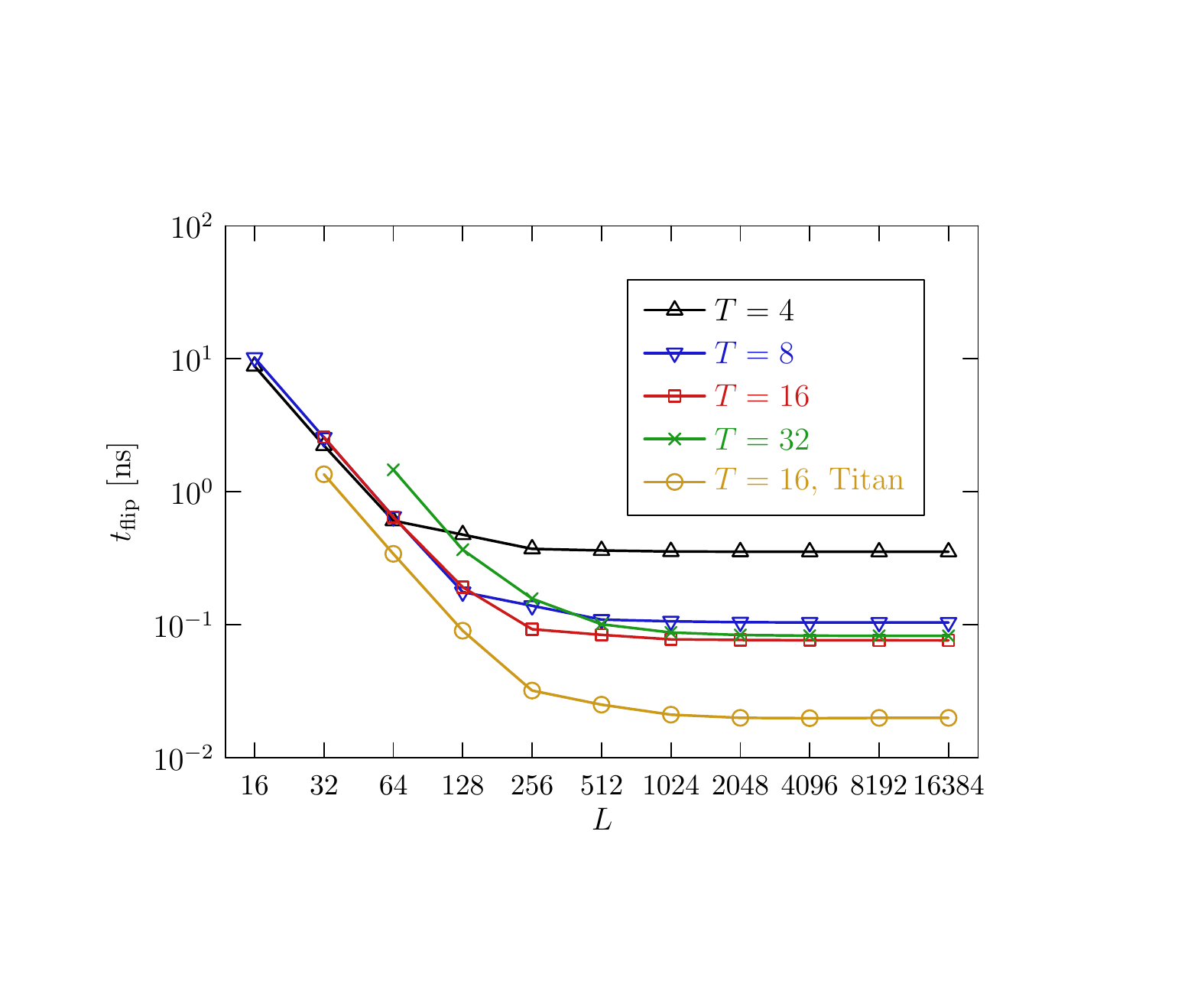}
  \caption
  { Spin-flip times in ns of the double-checkerboard 2D Ising model GPU simulation
    code as a function of linear system size $L$ and for tiles of size $T\times T$
    spins. The last data set is for runs on the GTX Titan Black GPU (Maxwell
    generation), whereas all other data is for the C1060 GPU (Tesla generation).  }
  \label{fig:blocks}
\end{figure}

While for $k=1$, the double checkerboard version of the code is slightly slower than
the optimized variant not relying on shared memory ($t_\mathrm{flip}=0.081$~ns on the
Titan Black, reducing the speedup compared to the scalar CPU code to 75), for $k>1$
one finds significantly improved performance yielding, for instance,
$t_\mathrm{flip}=0.020$~ns for $k=100$ (again for $L=4096$). This is comparable to
the results achievable with multi-spin coding \cite{lulli:15}. We note that the
tiling introduces as an additional parameter the tile size $T$ which is limited by
the maximum allowed number of threads per block (1024 on current devices, 512 for the
Tesla C1060). The dependence of spin-flip times on tile size is illustrated for the
C1060 GPU in Fig.~\ref{fig:blocks}. For more recent devices one finds the same
trend. In general it is preferable to have larger blocks as this helps to maximize
the number of resident threads per multiprocessor \index{occupancy} (occupancy)
which, in turn, generally improves the efficiency of the latency hiding
mechanism. The strong dependence of spin-flip times on the lattice size $L$ visible
in Fig.~\ref{fig:blocks} shows that for this type of problem optimal performance is
only achieved for rather large lattices. For more recent GPUs which feature roughly
10 times more cores than the C1060, this effect is even more pronounced as is
illustrated by the additional data for the GTX Titan Black GPU also shown in
Fig.~\ref{fig:blocks}. The simulation of a single, small lattice system just does not
provide enough parallelism to saturate the available parallel compute units ---
observations of this type led Gustafson to introduce the \index{scaling!weak}
weak-scaling scenario as discussed in Sec.~\ref{sec:scaling}. As we shall see below,
GPU codes for disordered systems or using generalized-ensemble simulation methods
such as multicanonical or population annealing simulations do not have this problem
and are able to fully load GPUs already for the smallest system sizes.

\subsection{Random-hit algorithms}

\index{random-hit~updates|(}
While the checkerboard update discussed in the previous section has the same
stationary distribution as the random-site or sequential-update schemes, the dynamics
of the different algorithms are not the same. Ideally, one would thus like to
implement the physically most plausible random-site selection algorithm, but
parallelizing it is a challenge as each single step typically is quite light
computationally and so does not provide enough work for parallelization. If
quantitative details of the dynamics are not of interest and the focus is on
universal properties, for example in studies of domain growth \cite{puri:09}, it can
be sufficient to concentrate on the checkerboard (or stochastic cellular automaton
\cite{wolfram:02}) dynamics, which is typically closer to the behavior of the
random-site algorithm than the sequential approach. If this is not sufficient or a
time resolution of less than a full sweep is required, updates based on the standard
checkerboard scheme are not suitable and, instead, a number of strategies for
parallelizing the random-site selection update can be employed:
\begin{itemize}
\item Single-site updates are independent of each other, and can hence be implemented
  in parallel, if they occur further apart than the range of interactions. In a
  domain decomposition of the lattice (tiling), this can be guaranteed by excluding
  the sites at the boundary of each tile from update attempts ({\em dead border\/}
  scheme) \cite{kelling:11}. To allow these border sites to be updated as well and
  thus make the algorithm ergodic, the origin of the tiling is randomly shifted to a
  different location in periodic intervals. The freezing of boundaries for certain
  time periods leads to weak dynamical artifacts that reduce as the size of tiles is
  increased and also as times in between synchronization are decreased
  \cite{kelling:17}. Another approach uses a sub-division of each tile into patches
  that touch only two of the neighboring tiles (for example by dividing a square tile
  into four equal sub-tiles). The sub-tiles are then updated in a random order which
  is the same for all tiles, such that updates never interfere with active sub-tiles
  in other tiles. For a surface growth problem, this {\em double tiling} approach was
  found to show weaker artifacts as compared to the scalar random-site update than
  the dead-border schemes \cite{kelling:17}. The random site selection in different
  tiles can only be implemented without massive penalties from non-coalesced memory
  transactions if the tiles are buffered in shared memory which, as for the multi-hit
  checkerboard scheme discussed above, is most efficient if the tiles receive several
  updates before synchronization.
  \index{n-fold~way|(}
\item At low temperatures, event driven simulations such as the $n$-fold way
  \cite{bortz:75} and the waiting time method \cite{dall:01} promise significant
  speedups as compared to standard Metropolis or heatbath methods, which produce many
  rejections for large $\beta$ due to the form of the probabilities in
  Eqs.~(\ref{eq:metropolis}) and (\ref{eq:heatbath}). If these are applied in
  parallel using a domain decomposition, the synchronization problem shows up in
  asynchronous clocks in different domains. To avoid parallelization bias a flip
  attempt on the boundary of a given tile can only proceed if the local time of the
  tile is not ahead of that of any neighboring tile \cite{lubachevsky:88}. Depending
  on the model, the profile of local times of the tiles can show roughening, thus
  destroying the scaling properties of the parallel implementation as more and more
  tiles need to idle until the times of their neighbors have advanced sufficiently
  \cite{korniss:00}. Possible remedies can be the introduction of long-range
  interactions \cite{korniss:03} or other approaches \cite{shchur:04}, but we will
  not discuss these further here.
  \index{n-fold~way|)}
\item Another strategy for parallelizing the random-site update that completely
  avoids approximations from domain decompositions is in simulating several copies of
  the system in parallel. If one is interested in the relaxation to equilibrium (for
  example for studying coarsening and domain growth \cite{puri:09}), this approach is
  well suited. If one wants to sample the dynamics {\em in equilibrium\/}, it has the
  downside of multiplying the total work spent on equilibration by the number of
  copies simulated in parallel. Depending on the system sizes and time scales of
  interest, it might be useful to choose a hybrid approach, where some parallelism is
  used to simulate different copies and some is used to speed up the updating of each
  copy using domain decompositions and the approaches discussed above
  \cite{kelling:17a}. An important consideration of simulating several systems in
  parallel with random-site updates is memory locality: if each system independently
  chooses the next site to update at random, memory accesses will be scattered,
  leading to poor performance. Much better performance is achieved if using the same
  sequence of random numbers for site selection in all copies and storing the
  configurations such that spins at the same lattice site but in different copies
  appear next to each other in memory. This is not a problem for updates that involve
  additional randomness at the site-updating step, as is the case for the Metropolis
  and heatbath updates of the Ising model, but would lead to identical trajectories
  for cases where the site selection is the only random step \cite{kelling:17}.
\end{itemize}
Using a combination of the above techniques, excellent GPU performance comparable to
that of the checkerboard approach can be achieved also for simulations with
random-site updates
\index{random-hit~updates|)}

\subsection{Cluster updates}
\label{sec:cluster}

\index{cluster~algorithm|(} While local spin updates of the Metropolis or heatbath
type in general work well and, as shown above, can be efficiently parallelized, it is
well known that in the vicinity of continuous phase transitions
\index{phase~transition} they are affected by an increased correlation between
successive configurations known as {\em critical slowing down\/}
\cite{binder:book2}. It is intuitively understood by the build-up of correlations
near criticality, with the ensuing increase in correlation length $\xi$ resulting in
a corresponding increase in the number of steps $\tau$ required to decorrelate
configurations. The assumption that decorrelation for local updates is based on the
diffusive propagation of information through the configurations leads one to expect a
relation $\tau \sim \xi^z$ with $z \approx 2$ \cite{sokal:97}. An effective antidote
against critical slowing down consists of cluster algorithms constructed in such a
way that they update typical clusters of size $\xi$ in one step. The identification
of such objects is difficult in general, and it has only been achieved in the full
sense for a number of spin models. For the Potts model, the relevant clusters are
those described by the Fortuin-Kasteleyn representation \cite{fortuin:72a}. The
corresponding update is implemented in the Swendsen-Wang cluster algorithm
\cite{swendsen-wang:87a} and its single-cluster variant proposed by Wolff in
Ref.~[\refcite{wolff:89a}], where he also introduced a generalization to
continuous-spin models. A number of further generalizations, including algorithms for
particle systems, have been proposed \cite{luijten:06,flores:17}.

For the Ising model (\ref{eq:ising_model}), the Swendsen-Wang update involves the
following steps:
\begin{enumerate}
\item For a given spin configuration define bond variables $n_{ij}$. For
  antiparallel spins, $s_i \ne s_j$, set $n_{ij} = 0$. For parallel spins, $s_i =
  s_j$, draw a random number $r$ uniformly in $[0,1)$ and set
  \[
    n_{ij} = \left\{
      \begin{array}{l@{\hspace{0.5cm}}l}
        1, & \mbox{if}\;r < p = 1-\exp(-2\beta), \\
        0, & \mbox{otherwise}.
      \end{array}
    \right.
  \]
\item Identify the connected components (clusters) of the lattice induced by the bond variables
  $n_{ij}$.
\item Flip each cluster of spins independently with probability $1/2$.
\end{enumerate}
Parallel implementations of steps (1) and (3) are quite straightforward as they are
completely local procedures, but the cluster identification in step (2) needs to deal
with structures that potentially span the whole system as the Fortuin-Kasteleyn
clusters used here undergo a \index{percolation} percolation transition
\index{phase~transition} right at the thermal critical point of the model
\cite{fortuin:72a}. An number of parallelization strategies for this step were
previously discussed for instance in
Refs.~[\refcite{heermann:90,baillie:91,flanigan:95}]. Note that connected-component
identification \index{connected~components} is of relevance in computer vision such
that significant effort is still being devoted to the development of ever more
efficient (serial and parallel) implementations of this algorithm \cite{he:17}.

The bond activation step is straightforwardly parallelized by assigning one thread to
each spin (or, possibly, one thread to a small tile of a few spins) and letting each
thread decide about the values of the two bond variables $n_{ij}$ connecting a spin
to neighbors with larger site indices (generalizations to other lattices are
straightforward, and each thread then deals with $z/2$ bonds, where $z$ is the
coordination number). The $n_{ij}$ represent one bit of information, and in the
interest of saving bus bandwidth, it makes sense to store them in the narrowest
native variables available (typically 8-bit wide integers) or possibly to use
``multi-bond coding'' to merge the states of several bonds into one word. It is again
advisable to use inline random-number generators, one per thread, for deciding about
$n_{ij}$ in case of parallel spins. The tiles covered by each thread block are best
chosen in the form of long strips as this increases the proportion of coalesced
memory transactions \cite{weigel:10b}.

\begin{figure}[tb]
  \centering
  \includegraphics[clip=true,keepaspectratio=true,width=0.4\columnwidth]{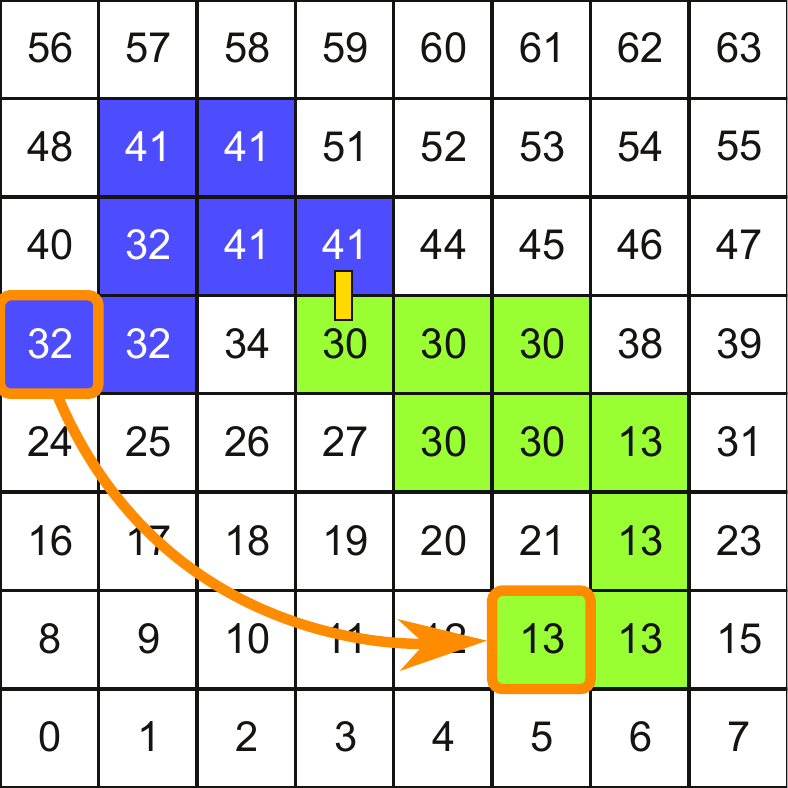}\hspace*{0.5cm}
  \includegraphics[clip=true,keepaspectratio=true,width=0.4\columnwidth]{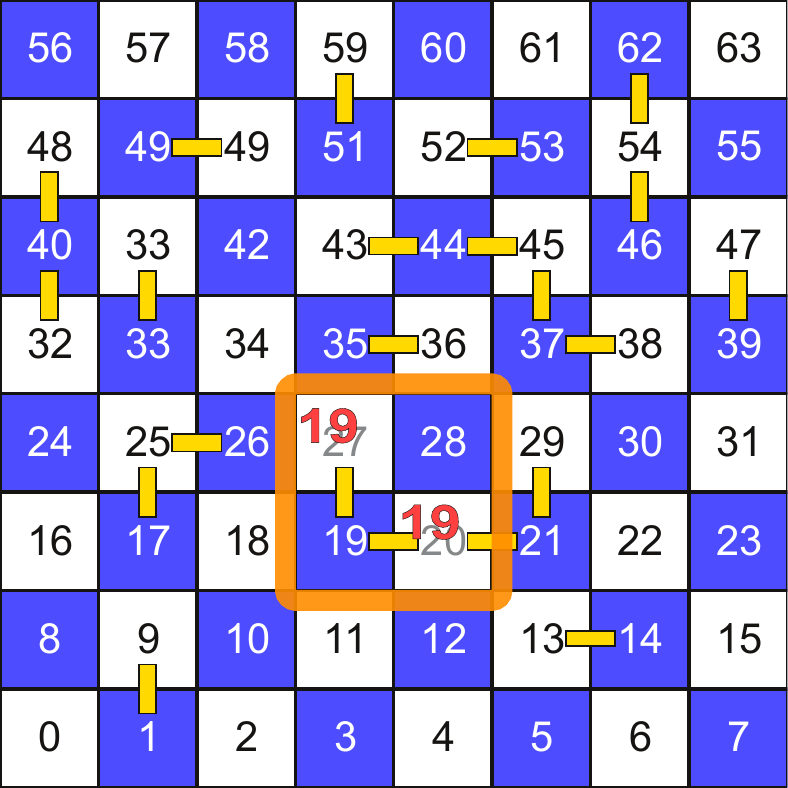}
  \caption
  {Left: Tree-based union-and-find algorithm applied to the connected component
    labeling. The edge between sites 30 and 41 is inserted, leading to an attachment
    of the smaller cluster tree to the root of the larger cluster. Right: Cluster
    identification by the self-labeling algorithm, using one thread for a tile of
    $2\times 2$ spins.}
  \label{fig:cluster-labeling}
\end{figure}

\index{connected~components|(}

The cluster-identification step is somewhat more intricate to parallelize. A number
of parallel cluster labeling techniques is discussed and compared in
Ref.~[\refcite{weigel:10b}]. The strategy taken there is to use a
\index{domain~decomposition} domain decomposition (tiling) to create a correct
cluster labeling in the tiles, i.e., ignoring couplings that cross the tile
boundaries, to then, in a second phase of the algorithm, consolidate cluster labels
across tile boundaries. Most of the available labeling algorithms are variants of the
following three approaches \cite{weigel:10b,he:17}:
\begin{enumerate}
\item Breadth-first or depth first searches, sometimes called ``ants in the
  labyrinth'' or label propagation, proceed from a seed site for each cluster and
  iteratively add neighbors of already discovered cluster sites to the cluster, thus
  ``growing'' it until no more connected unlabeled sites are found. The breadth-first
  strategy leads to an onion-shell structure, where the time step at which a given
  site is discovered is given by the chemical distance (shortest path) of the site to
  the seed. While these approaches are very intuitive and also efficient for serial
  implementations, they are not very suitable for parallel computing. Parallel work
  arises in the breadth-first variant in the wave front of growth sites (the current
  onion layer during growth). However, this parallelism is rather moderate and also
  \index{parallelism!irregular} irregular in that the number of wave-front sites
  fluctuates strongly.  The total work of these algorithms scales linearly with the
  number of sites.
\item Union-and-find or label equivalence algorithms provide solutions for the
  general problem of finding equivalence classes of nodes (i.e., connected
  components) under the sequential insertion of edges. A special case is the
  Hoshen-Kopelman algorithm well known for percolation applications
  \cite{hoshen:79}. A classical tree-based approach due to Tarjan \cite{tarjan:75}
  represents clusters as trees with a unique root site. The {\em find\/} operation
  for a given site corresponds to an upward tree traversal to find the root site that
  holds the cluster label. On the insertion of an external edge that merges two trees
  ({\em union\/} step), the smaller tree is attached to the root of the larger
  one. This prescription, called {\em balancing\/}, is chosen to result in flatter
  average tree heights and hence quicker find operations revealing the roots
  \cite{newman:01a}. If additionally each find step redirects the pointers of
  intermediate visited sites directly to the root ({\em path compression\/}), it can
  be shown that the algorithm performs both operations, union and find, in
  effectively constant time, such that the total computational effort is again linear
  in the number of lattice sites \cite{tarjan:75,newman:01a,elci:13}. The basic
  procedure is illustrated in the left panel of Fig.~\ref{fig:cluster-labeling}. In
  this full version, the approach is not well parallelizable as the tree
  manipulations need to occur in a well defined order and avoiding
  \index{race~condition} races in order not to result in corruption of the forest
  data structure.
\item A much more regular, iterative approach, sometimes called self-labeling
  \cite{baillie:91,weigel:10b}, is illustrated in the right panel of
  Fig.~\ref{fig:cluster-labeling}. Here the labels of the neighbors of each site are
  inspected and all of them are set to the minimum of the labels encountered. To
  result in a correct labeling on the whole tile, this procedure must be repeated
  until no label is changed during a full iteration. The number of iterations
  required is related to the length of the shortest paths connecting any two points
  on the same cluster. For a critical spin model on a tile of edge length $T$, this
  is known to scale as $T^{d_\mathrm{min}}$, where $1 < d_\mathrm{min} < d$ denotes
  the shortest-path fractal dimension of the model \cite{stauffer:book}. Hence the
  total work to achieve a full labeling with this approach scales as
  $T^{d+d_\mathrm{min}}$ as compared to the scaling proportional to $T^d$ for the
  label propagation and label equivalence methods. While it is hence asymptotically
  more expensive, the advantage of the self-labeling approach lies in the efficient
  parallelization as the label minimization in neighborhoods can be performed for all
  spins in parallel. This clearly leads to \index{race~condition} race conditions
  between threads in reading and writing the updated labels. These could be resolved
  using \index{atomic~operation} atomic operations,\footnote{Atomic operations on GPU
    are intrinsic instructions provided by CUDA that read, modify and write back
    memory locations with the guarantee that no other thread can perform a write
    after the read and before the write of the current thread \cite{cuda}.} but
  since the stopping condition is connected to an iteration without writes, the full
  sequence will converge faster to the same final label set without atomic
  operations.
\end{enumerate}

\index{connected~components|)}

\begin{figure}[tb]
  \centering
  \raisebox{-0.07cm}{\includegraphics[clip=true,keepaspectratio=true,width=0.411\columnwidth]{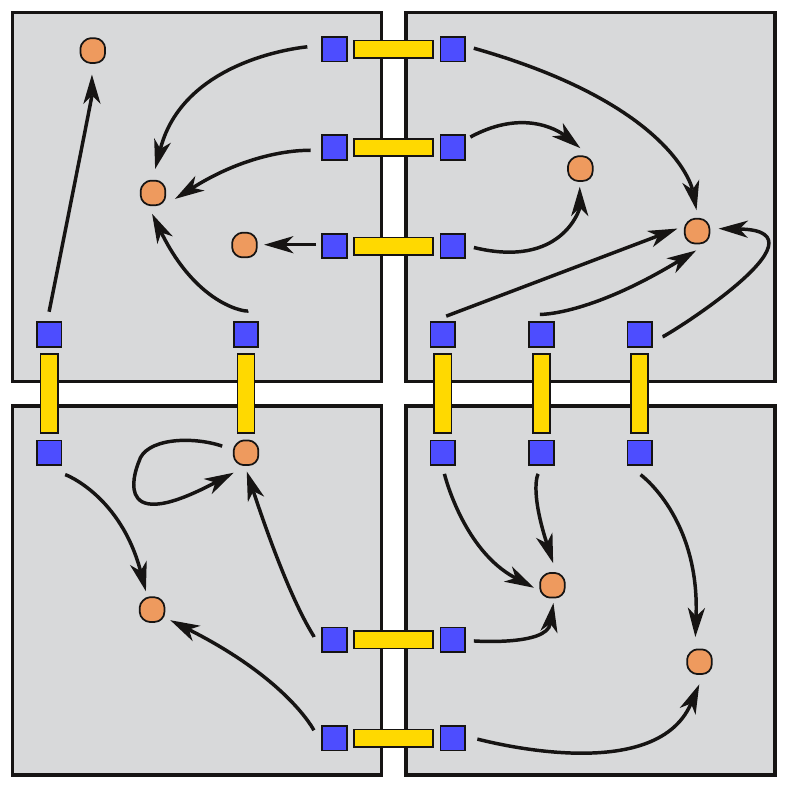}}\hspace*{0.5cm}
  \includegraphics[clip=true,keepaspectratio=true,width=0.4\columnwidth]{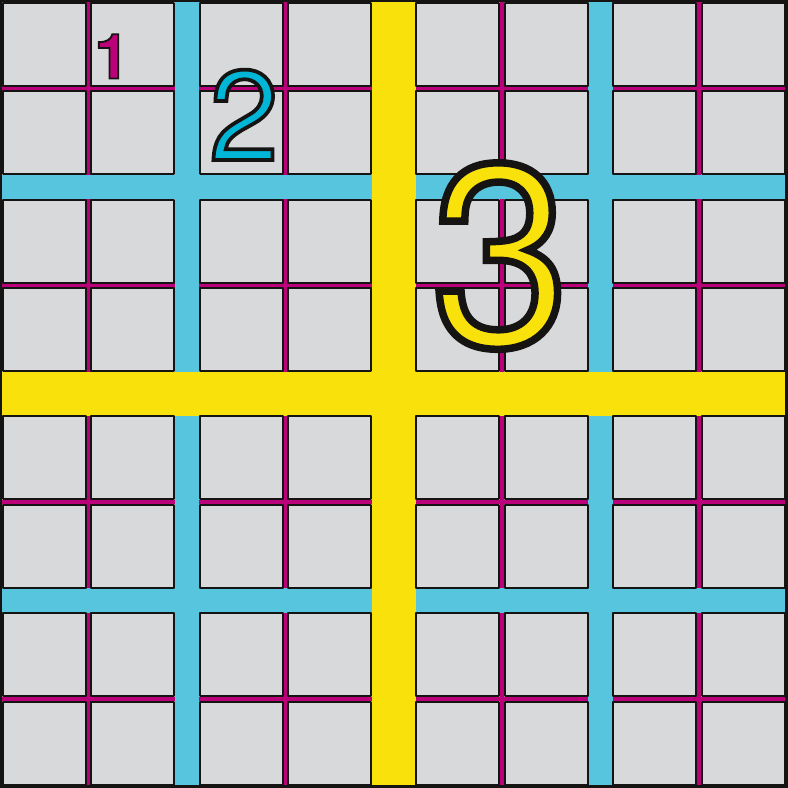}
  \caption{Left: Label consolidation between tiles using a relaxation
    mechanism. Neighboring tiles exchange the information about root sites (circles)
    for sites connected by bonds crossing tile boundaries (squares) and attach the
    smaller cluster to the larger one as in the balanced union-and-find
    approach. Right: Hierarchical sewing of tiles, where the missing edges are
    inserted at levels of increasing length scales leading to a combination of
    $2\times 2$ tiles of each level. In the example, there are 64 level-1 tiles, 16
    level-2 tiles and 4 level-3 tiles.}
  \label{fig:consolidation}
\end{figure}

Once a labeling in tiles has been achieved, the effect of the boundary bonds needs to
be taken into account. These link clusters in different tiles and hence the
corresponding cluster labels need to be identified. Since clusters might span several
tiles (especially close to the critical point where percolation first occurs), this
process might lead to a relabeling of a significant fraction of lattice sites. Two
possible algorithms to achieve this label consolidation are as follows:
\begin{enumerate}
\item In {\em label relaxation\/}, in a first step for each tile the information
  about root nodes for all boundary sites with active edges crossing the boundary is
  collected in an array. Then this information is exchanged with the neighboring tile
  and both root labels are set to the minimum of the two labels, thus merging the two
  clusters. Relaxation steps need to be repeated for the whole lattice until no
  further label updates occur. The corresponding setup is shown in the left panel of
  Fig.~\ref{fig:consolidation}. From a shortest-path argument similar to the one
  presented above for the self-labeling approach it is clear that at criticality the
  number of such iterations scales proportional to $\ell^{d_\mathrm{min}}$, where
  $\ell = L/T$ is the lattice size in units of tiles. For the 2D Ising model
  $d_\mathrm{min} = 1.08(1)$ \cite{miranda:91}.
\item The alternative {\em hierarchical sewing\/} method consists of adding the bonds
  that cross tile boundaries first only between sets of $2\times 2$ tiles, reducing
  the total number of tiles by a factor of four (level 1), then inserting the bonds
  between $2\times 2$ of the resulting larger tiles (level 2) etc., until all of the
  boundary bonds have been inserted. This is illustrated in the right panel of
  Fig.~\ref{fig:consolidation}. The approach corresponds to a divide-and-conquer
  strategy and the required number of steps is logarithmic in the effective lattice
  size $\ell$. The load of the device decreases at each level, and at some point some
  of the compute units will be idling, but this effect is found not to be very severe
  for the achievable system sizes.
\end{enumerate}
The required number of operations and parallel performance of this step can be
explicitly understood from simple calculations, and the numerical findings are
consistent with these considerations, for details see Ref.~[\refcite{weigel:10b}]. It
turns out that the hierarchical approach performs better than the relaxation
mechanism for large system sizes, but for intermediate sizes the relaxation approach
(leading to a better resource utilization) is the better choice.

Finally, the cluster flipping step is again quite straightforward to parallelize. New
spin orientations are drawn for all sites first and then each site finds its root and
sets its own orientation to that of the root site. The total speedups achievable with
this approach for simulations of the 2D Ising model are somewhat less impressive than
for the single-spin flip updates, but comparing the Fermi card GTX480 to an Intel
Q6700 CPU a system of size $L=8192$ can still achieve about a 30-fold speedup
\cite{weigel:10b}. A variant of the methods outlined here provides comparable
performance for larger systems, but better results for small sizes and somewhat
simpler code \cite{komura:12}: it uses self-labeling, but after each single iteration
a step of path compression (or label propagation) in the spirit of the union-and-find
approaches is performed first before moving on to the next phase of
self-labeling. Further improvements have been suggested, in particular related to
using \index{atomic~operation} atomic functions to perform the label consolidation in
full parallelism \cite{wende:13,komura:15}.

\index{cluster~algorithm|)}

\subsection{Continuous spins}

The considerations to this point have concentrated on systems with discrete degrees
of freedom, such that floating-point performance was not an issue. This changes for
systems with continuous spins such as the O($n$) model with Hamiltonian
\[
{\cal H} = -J\sum_{\langle i,j\rangle} \bm{s}_i\cdot\bm{s}_j,\;\;\;|\bm{s}_i| = 1,
\]
where the $\bm{s}_i$ are $n$-component vectors of units length. A particular feature
of GPU devices is that they are designed for single-precision arithmetics --- the
excess precision provided by double precision variables is mostly irrelevant for the
graphics calculations that are the initial target of GPUs. While early generations
did not provide double precision arithmetics at all and even single-precision
calculations were not fully compliant with the IEEE-754 floating-point standard, more
recent cards are IEEE compliant and a double precision mode is now available. It
remains true, however, that double-precision calculations are significantly slower
than single-precision ones, in particular on consumer cards which feature 4--5 times
less double-precision floating-point performance \index{floating-point~performance}
than the Tesla boards.\footnote{The most recent Pascal generation of Nvidia GPUs also
  has native support for half-precision (16 bit) floating-point calculations which
  are also much slower on the consumer cards than on the GPGPU boards.} It is hence
reasonable to carefully consider which level of precision is required for a
particular calculation. For the task of simulating continuous-spin models, the
following considerations should be taken into account:
\begin{itemize}
\item As an implementation of a stochastic process, numerical stability and accuracy
  are not as much of an issue for Monte Carlo simulations than for, say, molecular
  dynamics. Round-off error is not expected to accumulate and lead to systematic
  errors in calculations. It hence will often be acceptable to use single-precision
  floating-point variables to represent the spins and only use double precision for
  aggregate quantities such as energies, magnetizations and other measured
  observables. Numerical tests confirm that stability is not an issue
  \cite{weigel:10a}. When using a Cartesian representation of the spins $\bm{s}_i$
  and if updating the spins by adding modification vectors, it might be necessary to
  periodically renormalize the spin lengths to unity, similar to what is done on CPU
  also.
\item To the extent that calculations are memory bound, the use of narrower variables
  such as single-precision floats instead of doubles will also help reduce memory
  bandwidth pressure and hence additionally improve throughput over the mere increase
  in actual arithmetic performance. From this perspective, it can also be
  advantageous to use a polar representation of the spins, trading off memory
  bandwidth (and storage if that is a concern) against the typically higher
  arithmetic load in a spherical representation for performing calculations such as
  computing the interaction energy \cite{yavorskii:12}.
\index{floating-point~performance|(}
\item A further heritage from the 3D graphics world are special-function units
  allowing to perform some of the most common evaluations of transcendental
  functions, including sine, cosine, exponential and logarithm functions, in
  hardware. These provide particularly high performance, but at the expense of
  slightly reduced precision as compared to the IEEE-754 standard (typically a few
  units in the least-significant digit) \cite{hwu:11}. Replacing evaluations of
  exponential functions, for example in the Metropolis criterion, of the logarithms
  occurring, for instance, in the Box-Muller algorithm for generating Gaussian random
  numbers, as well as of trigonometric functions for instance in scalar products by
  calls to the special function units available in GPU hardware can lead to
  significant speedups. While the small reduction in precision as compared to the
  library implementations is typically not a problem, one side effect is that such
  GPU codes will no longer fully agree in all bits of output with the corresponding
  CPU codes. Small differences in the value of the exponential in the Metropolis
  criterion, in particular, can lead to a divergence of Monte Carlo trajectories
  \cite{barash:16}. To the extent that the output also depends on the order of
  operations for evaluating sums, updating spins etc., full identity of outputs
  between GPU and CPU codes cannot be expected anyway, however.
\end{itemize}
\index{floating-point~performance|)}

\begin{table}[tb]
  \centering
  \tbl{Times per spin flip in ns for various implementations of single-spin flip
    simulations of the 2D Heisenberg model and speedup factors as compared to the CPU
    reference implementation. All data are for multi-hit updates with $k=100$ and
    system size $L=4096$.\label{tab:heisen}}
  {\begin{tabular}{llt{2}t{0}} \toprule
    \multicolumn{1}{c}{device} & \multicolumn{1}{c}{mode} & \multicolumn{1}{c}{$t_\mathrm{flip}$ [ns]} & \multicolumn{1}{c}{speedup} \\ \colrule
    CPU (Intel Xeon E5-2620 v3) & float  & 98.6   & 1     \\
                               & double & 186.5  & 1     \\
    Tesla C1060     & float         &  0.74                                 & 133   \\
    & float, fast\_math   &           0.30                        & 329   \\
    & double        &  4.66                                 & 40    \\ 
    Tesla K20m     & float         &  0.177                                 & 557   \\
    & float, fast\_math   &           0.149                        & 662   \\
    & double        &  0.408                                 & 457    \\ 
    GTX Titan Black & float         &   0.152                                & 649   \\
    & float, fast\_math   &  0.105                                      & 939   \\
    & double        &                 0.323                  & 578    \\ \botrule
  \end{tabular}}
\end{table}

\index{floating-point~performance|(}
For simulations of the Heisenberg model on the square lattice, one finds quite
significant performance differences, see the data collected in Table
\ref{tab:heisen}. For the reference CPU implementation we find about a factor of two
difference between the single-precision and double-precision variants of the code. We
note that this is probably mostly a cache effect, however, since for smaller system
sizes, when the configuration fully fits into the cache, the difference in
performance between single and double precision is negligible (and it corresponds to
the 98 ns performance shown in the table for single precision). For the
first-generation Tesla C1060 card, the difference between single and double precision
performance was dramatic, even taking into account the cache effect seen on CPU. For
the more recent cards, the performance in single and double precision is much more
comparable, and in fact some very large speedup factors are observed there. Usage of
the fast special function intrinsics provides an additional 15-30\% improvement on
these cards. We note that although the Titan Black is a consumer card, it has
significantly higher double-precision performance than other consumer cards, so we do
not see the possibly expected performance degradation here as compared to the K20m.
\index{floating-point~performance|)}

\section{Random number generation}
\label{sec:RNG}

\index{random~number~generator|(}

Although stochastic algorithms formally depend on the input of a stream of random
bits to model the noise, such as the random acceptance of spin flips in the
Metropolis algorithm, it is normally not feasible to use a true source of randomness
(such as, e.g., a quantum mechanical system) for this purpose --- the rate at which
random numbers are consumed by Monte Carlo simulations of systems with simple energy
functions, $10^6$ to $10^9$ per second on current computers, is too large to make
this feasible.\footnote{Note, however, that recently optical methods allow to generate
  true random numbers at rates of several hundred MBit/s \cite{qi:10}, such that the
  use of such physical randomness in simulations might become feasible in the near
  future.} Instead, simulations normally rely on the use of pseudo-random number
generators that are based on deterministic arithmetic relations
\cite{gentle:03}. Such schemes need to balance desirable implementation properties
such as efficiency, reproducability, portability etc.\ with the most fundamental need
of delivering high-quality random numbers. While the output of a deterministic
algorithm can never pass as a truly random sequence since the algorithm itself allows
to predict the next number with certainty, a number of general statistical tests such
as the equi-distribution of $n$-tuples of numbers in $n$ dimensions are routinely
used to assess the quality of a given generator. The current de facto standard in
this respect is the test battery ``TestU01'' developed by L'Ecuyer and co-workers
\cite{lecuyer:07}. It is good practice to ensure that generators used for production
runs and, in particular, high-precision studies pass the tests in such suites. Even
for generators passing such tests, however, it is possible that subtle correlations
in pseudo-random sequences interact with particular properties of a model of
statistical physics and the algorithm used to simulate it in such as way as to
produce large biases \cite{ferrenberg:92}. It is hence often advisable to compare the
results from simulations using identical algorithms and implementations, but random
sequences from different generators to exclude such problems.

Random number generators (RNGs) for massively parallel simulations need to satisfy
additional demands that are not (so) relevant in serial calculations. There is a
large number of threads consuming random numbers in simulations of the type discussed
here, so in order to avoid a bottleneck in parallel scaling one needs the same or at
least a similar number of threads for generating these random numbers. This can be in
a separate kernel such that the produced numbers are stored in an array that is later
on consumed by individual threads in the simulation kernel, or through an in-lining
of individual RNG instances in the simulation kernel itself
\cite{manssen:12,curand}. We have typically found the latter approach to be
advantageous as it is more flexible and reduces the memory requirements. In both
cases, it is crucial to ensure that the sub-sequences of random numbers finally used
by individual threads are sufficiently uncorrelated. This might be achieved (a)
through a seeding or parameter choice of individual generator instances that ensure
such a lack of correlation, (b) through a partitioning of the same global sequence
between the individual threads that lead to uncorrelated sub-sequences, or (c)
through the use of generators with extremely large periods such that a random choice
of sub-sequences has a sufficiently small probability of overlap with other
sequences. If skipping is used to ensure that non-overlapping sub-sequences are
assigned to different threads, the fact that the random numbers are used in a
different than the sequential order in which they are usually fed into the test
suites can lead to quite different quality results. A second important consideration
concerns the memory footprint of the state of the generator. As the states need to be
transferred to and from the compute units for every (possibly multi-hit) update, the
corresponding transfers can easily turn into performance limiting factors for
generators with larger states. Also, to achieve good performance for the calculations
required for generating random numbers one might want to store the states in shared
memory which often means that only a few bytes per thread are available.

\begin{table}[tb]
  \tbl{Properties of some random-number generators implemented on GPU. All
    random-number sequences were fed through the tests of the TestU01 suite \cite{lecuyer:07}
    which is divided into the ``SmallCrush'', ``Crush'' and ``BigCrush'' sections of
    increasing stringency. If at a given stage too many failures occurred, the more
    advanced tests were not attempted.  The performance column shows the peak number
    of 32-bit uniform floating-point random numbers produced per second on a fully loaded GTX~480
    device. Note that the Philox generators, albeit
    occupying local memory of $4\,\times\,32$ bits for number generation, do not require to
    transfer a ``state'' from and to global memory as long as the generator keys are
    deduced from intrinsic variables such as particle numbers etc.
    \label{tab:generators}}
  {\begin{tabular}{l@{\hspace{0.2cm}}c@{\hspace{0.15cm}}c@{\hspace{0.15cm}}c@{\hspace{0.15cm}}ccl} \toprule
     \multicolumn{1}{c}{generator} & bits/thread & \multicolumn{3}{c}{failures in TestU01} & Ising test & perf. \\
                                   &                  & SmallCrush & Crush  & BigCrush        &   & $\times 10^9$/s \\ \colrule
     LCG32     &               32 &      12    &   ---  &     ---         & passed & 58 \\
     LCG32, random &           32 &       3    & 14     &     ---         & passed  & 58 \\
     LCG64     &               64 &       None &  6     &     ---         & failed  & 46 \\
     LCG64, random &           64 &       None &  2     &     8           & passed  & 46 \\
     MWC       &          64 + 32 &       1    & 29     &     ---         & passed  & 44 \\
     Fibonacci, $r=521$ & $\ge 80$ &      None & 2      &     ---         & failed  & 23 \\
     Fibonacci, $r=1279$ & $\ge 80$ &     None & (1)    &     2           & passed  & 23 \\
     XORWOW (cuRAND) &          192 &       None & None   & 1/3             & failed & 19 \\
     MTGP (cuRAND) &       $\ge 44$ &       None &    2   &    2            & ---     & 18 \\
     XORShift/Weyl &           32 &       None & None   & None            & passed  & 18 \\
     Philox4x32\_7 &       (128)  &       None & None   & None            & passed  & 41 \\
     Philox4x32\_10 &      (128)  &       None & None   & None            & passed  & 30 \\ \botrule
   \end{tabular}
 }
\end{table}

In the following, we summarize the properties and suitability of some of the most
common generators for GPU calculations. More details can be found in
Ref.~[\refcite{manssen:12} (see also Ref.~\refcite{barash:14}]). Each generator was
implemented on GPU and fed through the TestU01 battery of tests as well as in an
application test in simulating the 2D square-lattice Ising ferromagnet using the
Metropolis algorithm, where the results were compared against the exact expressions
for the finite-lattice energy and specific heat \cite{ferdinand:69a}. Performance
results were collected for the bare random-number generation as well as for the Ising
simulation test. The corresponding data are summarized in Table \ref{tab:generators}.
\begin{itemize}
\item {\bf Linear congruential generators (LCGs).} This most basic  class of generators
  follows the linear recursion
  \[
  x_{n+1} = ax_n+c \pmod{m},
  \]
  where $a$ and $c$ are integer constants and the modulus $m$ defines the total range
  of the numbers and hence the number of random bits generated in one call. As in
  most other generators the recursion is implemented in integer arithmetic, and the
  typically required floating-point numbers uniformly distributed in $[0,1)$ are
  generated via a suitable output function, here $u_n = x_n/m$. For good choices of
  $a$ and $c$ the period of the LCG is $p = m$, and if a native type is used for
  $x_n$ one is restricted to $m<2^{64}$, leading to rather modest period lengths,
  especially if taking into account that typically no more than $\sqrt{p}$ numbers
  should be used \cite{knuth:vol2}. It is straightforward to skip ahead in the
  sequence by an arbitrary number of steps with no additional computational effort
  \cite{gentle:03}, such that it is possible to concurrently generate non-overlapping
  sub-sequences by different threads in a simulation code. This, however, reduces the
  available period per thread even further. In the statistical and Ising application
  tests these methods, implemented for $m=2^{32}$ and $m=2^{64}$ with suitable
  constants $a$ and $c$ \cite{manssen:12}, yield poor results with the $32$-bit
  version even failing SmallCrush, see the data collected in Table
  \ref{tab:generators}. Somewhat better results are found if initializing each
  parallel generator instance with a random seed produced by a different LCG, thus
  introducing some additional randomness at the expense of no longer being able to
  guarantee non-overlapping sub-sequences. These generators are mostly useful due to
  their extremely simple and arithmetically light implementation and the very small
  state of 32 or 64 bits per threads, respectively, leading to the highest peak
  performances of generators discussed here (cf.\ Table~\ref{tab:generators}) and for
  testing purposes. The randomized 64-bit variant might be acceptable in some simple
  or lower precision applications as it passes the application test and most of the
  tests in the Crush battery.
\item {\bf Multiply with carry (MWC).} A generalization of the LCG approach due to
  Marsaglia \cite{marsaglia:91} is given by the recursion
  \[
  \begin{split}
    x_{n+1} &= ax_n +c_n \pmod{m},\\
    c_{n+1} &= \lfloor (ax_n+c_n)/m \rfloor.
  \end{split}
  \]
  In other words, the additive term $c_n$ in the $(n+1)$st step is the {\em carry\/}
  from the previous iteration, hence the name multiply-with-carry. If again using 32
  bits for the state vector, $x_n$ and $c_n$ together occupy $64$ bits. It is
  possible to generate a large number of ``good'' multipliers $c$ such that generator
  instances used by different threads use different values of $c$. MWC can be
  efficiently implemented on GPU \cite{ferrero:11}. The statistical quality of this
  generator turns out to be only marginally better than that for the pure $32$-bit
  LCG, see the data in Table~\ref{tab:generators}, while the storage requirement is
  64 bit for the state and an additional 32 bits for the multiplier $c$, such that
  the additional effort as compared to the LCGs does not seem to be worthwhile.
\item {\bf Lagged Fibonacci generators.} In the simplest case, these are based on a
  two-term lagged Fibonacci sequence with recursion
  \[
  x_{n} = a_s x_{n-s} \otimes a_r x_{n-r} \pmod{m},
  \]
  where the operator $\otimes$ typically denotes one of the four operations addition
  $+$, subtraction $-$, multiplication $\ast$ and bitwise XOR $\oplus$,
  respectively. For $32$-bit variables $x_n$, the state size is $32r$ bits, and it is
  then convenient to choose $m=2^{32}$. For $\oplus = +$ the maximal period is
  $2^{31}(2^r-1)$, which becomes very large for typical values of $r$. The generator
  can be conveniently implemented directly in floating-point arithmetic. For good
  quality one needs $r\gtrsim 100$, leading to relatively large storage requirements,
  but here the generation of $s$ random numbers can be vectorized by the $n$ threads
  of a block. Hence, the ring buffer of length $r+s$ 32-bit words is shared among the
  threads of a block, leading to a state size of $(r+s)/n$ words per thread. If $s$
  is chosen to be close to the number of threads in a block the total state is only a
  few words per thread \cite{weigel:10a}. Using the combinations $r=521$, $s=353$ and
  $r=1279$, $s=861$, respectively, and using random initial seeds per thread that
  should be safe thanks to the large overall period, we find reasonably good
  statistical results at least for the generator with larger $r$ and good
  performance, while the generator with $r=521$ leads to a failed Ising application
  test, see the data in Table~\ref{tab:generators}.
\item {\bf Mersenne twister.} This very popular generator in serial applications is
  also based on a generalized two-term lagged Fibonacci sequence \cite{matsumoto:98},
  where the larger lag $N$ is derived from a Mersenne prime $2^k-1$ as
  $N=\lceil k/32\rceil$, hence the name. To improve the equidistribution properties,
  the resulting sequence $x_n$ is additionally subjected to a tempering
  transformation, for details see Ref.~[\refcite{matsumoto:98}]. The maximal
  achievable period is $2^k-1$ which for the most popular choice $k=19937$ amounts to
  approximately $4\times 10^{6001}$. Similar to the case of the more standard lagged
  Fibonacci generators, Mersenne twister can only efficiently be used on GPU if the
  state is shared between different threads as otherwise the storage requirements are
  too large. A variant of the generator adapted for GPU was proposed in
  Ref.~[\refcite{saito:10}] under the name MTGP. It uses $k=11213$ and 256 threads to
  generate $N-M$ numbers in parallel, where $M < 95$ is the smaller lag. Since
  $N=\lceil 11213/32\rceil = 351$, the storage requirements per thread are
  $351/256 < 2$ words. An implementation of this algorithm is now part of the CURAND
  library that comes with Nvidia's CUDA distribution \cite{curand}. Independent
  sequences can be generated from 200 parameter sets resulting from number theoretic
  calculations \cite{saito:10} that are provided with the implementation. This yields
  a rather limited number of independent sequences, however, so that in practice
  possibly correlated sub-sequences from different initial seeds would need to be
  used. The fixed number of 256 threads required for MTGP also restricts the in-line
  use of this generator in the simulation kernel, and it is more suitable for
  pre-generating random numbers in a separate kernel to be stored in an array. As the
  data in Table \ref{tab:generators} demonstrate, the statistical quality of the
  generator is good, but it fails some tests based on $\mathbb{F}_2$ linearity. The
  generator speed is good, but not outstanding.
\item {\bf XORShift generators.} Another class of generators proposed by Marsaglia is
  based on the exclusive-or or bitwise sum operation applied to a state and a
  bit-shifted copy, known as XORShift \cite{marsaglia:03a}. The relevant recursion is
  given by
  \begin{equation}
    x_n = x_{n-1}(I \oplus L^a)
    (I \oplus R^b)(I \oplus L^c) =: x_{n-1} M,
    \label{threeshift}
  \end{equation}
  where $L$ denotes a left-shift by one bit, $R$ the corresponding right shift, and
  $I$ is the identity bit-matrix. For an appropriate choice of $a$, $b$ and $c$, the
  period is $2^w-1$, where $w$ is the word size. While the generators proposed in
  Ref.~[\refcite{marsaglia:03a}] used $w$ between $32$ and $192$ with relatively
  moderate properties, in Ref.~[\refcite{manssen:12}] a generator with $w=1024$ was
  proposed that is specifically optimized for GPU applications. There, the 32 threads
  of a warp share the 1024-bit state by contributing one 32-bit word each. The shifts
  and XORs are cooperatively implemented by the threads of a warp, while the state is
  stored as an array in shared memory. Since the threads in a warp operate in
  lockstep, no explicit synchronization \index{synchronization} is necessary. An
  additional tempering of the output sequence is achieved by combining the output of
  the XORShift with a Weyl sequence of the form $y_n = (y_{n-1}+c) \pmod{2^w}$ with
  an odd constant $c$ \cite{brent:07}. Full explanations and the source code of the
  CUDA implementation are provided in Ref.~[\refcite{manssen:12}]. Independent
  sub-sequences can be determined by using skip-ahead via the application of a
  precomputed power of the recursion matrix, and the provided implementation uses
  sub-sequences of length $w^{137} \approx 2\times 10^{41}$, which are safe from
  being exhausted on currently available hardware. On testing this generator it is
  found that all tests of the suite TestU01 and also the Ising application test are
  passed, no matter whether the sequence is used in single-thread or warp order. The
  performance of the initial implementation using shared memory is good, see Table
  \ref{tab:generators}; further improvements should be possible from using the thread
  shuffle \cite{cuda} instructions that allow to exchange data between the threads in
  a warp directly.
\item {\bf Counter-based generators.} A class of generators that are not derived from
  a recursion, but instead loosely based on secret-key cryptographic transformations
  was proposed in Ref.~[\refcite{salmon:11}]. Here, the idea is to generate the $n$th
  number in the sequence directly by applying some function $f$ to the index (or {\em
    counter\/}) $n$,
  \[
    x_n = f_k(n),
  \]
  such that knowledge of $x_n$ is not required to compute $x_{n+1}$. Suitable
  functions of this type can be derived from cryptographic codes such as DES and AES
  \cite{hellekalek:03}. If $n$ corresponds to the plaintext, it is clear that the
  ciphertext $f_k(n)$ must be statistically indistinguishable from a random sequence
  of bits for the code to be cryptographically secure. Standard codes ensuring this,
  such as DES and AES, are typically to slow, however, to serve as drop-in
  replacements for usual RNGs. Instead, Salmon {\em et al.\/} \cite{salmon:11}
  suggest to consider a simplified, ``mock'' version of AES with transformations
  based on integer division and its remainder,
  \[
    \begin{split}
      \operatorname{mulhi}(a,b) &= \lfloor (a\times b)/2^w \rfloor,\\
      \operatorname{mullo}(a,b) &= (a\times b) \mod 2^w,
    \end{split}
  \]
  such that the main iteration picks two consecutive words $(L,R)$ out of a block of
  $N$ words of $w$ bits each and computes
  \[
    \begin{split}
      L' &= \operatorname{mullo}(R,M),\\
      R' &= \operatorname{mulhi}(R,M)\oplus k \oplus L. 
    \end{split}
  \]
  The generator applies $r$ rounds of such transformations with different multipliers
  $M$, interspersed with additional permutations of the elements. This results in the
  generators dubbed Philox-Nxw\_r, where it is expected that a larger number of
  iterations $r$ improves the quality of the output, and $r\ge 7$ is suggested for
  good quality of the generated random numbers \cite{salmon:11}. This class of
  generators has two main advantages: (a) The state of $N\times w$ bits, where $N=4$
  and $w=32$ for one of the standard generators tested in
  Refs.~[\refcite{salmon:11,manssen:12}], can be arbitrarily split between the space
  of keys $k$ and the counter $n$ of random numbers in a given sequence. It is
  therefore straightforward to generate a large number of independent sub-sequences
  and, in particular, it might be advantageous to tie the sequence numbers to some
  intrinsic variables of the calculations such as the particle or spin number, the
  system size, the disorder realization etc., such that exactly the same random
  numbers are used independent of the actual distribution of work over the available
  compute units. (b) Since these generators are not based on a recursion, it is in
  fact not necessary to store and transfer state variables for generator instances
  used by individual threads. If the sub-sequences to be used are derived in a
  natural way from intrinsic variables such as the particle number etc., it suffices
  to pass an iteration number common to all threads into an updating kernel to have
  each thread generate the next number in the sequence. In addition to passing all
  tests of the suite TestU01 and the Ising application test, the Philox generators as
  implemented in Refs.~[\refcite{salmon:11,curand}] show excellent performance, see
  the data in Table~\ref{tab:generators}.
\end{itemize}
A number of further generators are made available in the program library of
Ref.~[\refcite{barash:14}].  The significant number of different generators might
appear confusing to the novice, and it is not in general necessary to make oneself
familiar with the details of all of them. Good general purpose generators with
excellent statistical properties and high performance are given by the XORShift
generator \cite{manssen:12} and the counter-based Philox \cite{salmon:11}. The latter
has the additional advantage of being available as part of Nvidia's CURAND library
\cite{curand}.

\index{random~number~generator|)}

\section{Generalized ensembles}
\label{sec:generalized}

While standard local-update Markov-chain algorithms such as the Metropolis and
heatbath methods discussed in Sec.~\ref{sec:canonical} are extremely general and
often also quite efficient, and cluster updates can be used for some systems in the
vicinity of critical points, there are a range of situations where these methods fail
to equilibrate the systems or do not give access to the quantities required in
certain contexts. This affects systems with \index{complex~free-energy~landscape}
rugged free-energy landscapes including but not limited to systems with quenched
disorder and certain biomolecular systems such as protein models \cite{janke:07}. For
such problems a range of generalized-ensemble simulation methods have become
available that allow to accelerate convergence by avoiding or overcoming barriers in
phase space.  Also, the sampling of rare events in systems with first-order
\index{phase~transition} phase transitions and methods giving access to the free
energy and microcanonical density-of-states call of special techniques, whose
suitability for massively parallel implementations will be discussed next.

\subsection{Parallel Tempering}

\index{parallel~tempering|(}

In parallel tempering or replica-exchange Monte Carlo a number $n_T$ of copies
(replicas) of the system are simulated in parallel, and each replica is held at a
different inverse temperature $\beta_i$ \cite{geyer:91,hukushima:96a}. The replicas
are labeled in order of increasing inverse temperatures, i.e., ensuring that
$\beta_i < \beta_j$ for $i < j$. At periodic intervals, an exchange of the
configurations $i$ and $j$ is proposed and accepted with the Metropolis probability
\begin{equation}
  p_\mathrm{acc}(i,j) = \min\left[1,e^{(\beta_i-\beta_j)(E_i-E_j)}\right],
  \label{eq:pt}
\end{equation}
where $E_i$ and $E_j$ are the configurational energies of the affected replicas.  It
is easy to see that this just satisfies the detailed balance condition (\ref{eq:DB})
with respect to the joint Boltzmann distribution of all replicas. In practice, one
normally only considers neighboring replicas, i.e., $j = i+1$, and attempts an
exchange of each pair of neighboring configurations. This scheme can greatly improve
relaxation as it allows replicas that are trapped in local minima at low temperatures
to escape to high temperatures where they can relax freely and later on return to low
temperatures, possibly occupying a different minimum. Ideally, copies hence perform a
random walk in temperature space. It is clear that non-negligible acceptance of such
exchange moves can only be expected if the typical energies of neighboring replicas
are comparable, that is if the energy histograms at temperatures $\beta_i$ and
$\beta_j$ have sufficient overlap \cite{janke:08}. Too few temperatures will hence
preclude the intended temperature random walks, but also too many temperatures are
not ideal as in a random walk the number of steps required to traverse the full
temperature range will grow as the square of the number of temperature points. One
hence expects an optimum number and distribution of temperature points, and possible
schemes for determining the optimum parameter set for parallel tempering simulations
have received a fair amount of attention
\cite{katzgraber:06,bittner:08,hasenbusch:10}.

Replica-exchange Monte Carlo appears to be an ideal match for parallel computing as
most of the time of a simulation will be spent on updating the replicas with some
conventional Monte Carlo algorithm (for example single spin flips), and the
occasional exchange of configurations according to Eq.~(\ref{eq:pt}) is very cheap
computationally, but also in terms of communications as no actual copying of
configurations is required. In a shared memory system, it is typically simplest to
exchange pointers to the configurations between neighboring replicas, whereas on a
distributed memory machine it is fastest to exchange the inverse temperatures
$\beta_i$ and $\beta_j$. In both cases, only a single integer or floating-point
variable needs to be communicated.\footnote{Note that in order to implement the latter
  efficiently, it is typically necessary to maintain {\em two\/} arrays, one mapping
  from replica indices to (inverse) temperatures and one mapping from (inverse)
  temperatures to replica indices as otherwise it is not possible to easily identify
  the replicas belonging to two neighboring temperatures for proposing an exchange
  move.}

Parallel tempering has been implemented on GPU for a range of systems, including spin
models \cite{weigel:10a}, polymers \cite{gross:11}, as well as spin glasses
\cite{fang:14,baity-jesi:14,lulli:15} and random field systems \cite{navarro:16}. In
terms of the work distribution, the actual replica-exchange step is so light that it
is typically irrelevant whether it is implemented on CPU or in a GPU kernel. Note,
however, that it requires up-to-date values for the configurational energies $E_i$.
The necessary calculations should either be done on-the-fly in the local-updating
kernel by adding the energy change incurred by the update of a degree of freedom to
the current value of the total energy (an approach that is particularly feasible if
the energy is an integer value such as for a discrete spin model), or distributed
over the GPU(s) via a dedicated energy-calculation kernel. We note that the typical
number of replicas, which is of the order of 10--100 for most applications, does not
provide enough parallelism to fully load a current GPU device. It is therefore
necessary to combine the parallelism provided by the replica-exchange algorithm with
further techniques such as a \index{domain~decomposition} domain decomposition
\cite{weigel:10a}, the trivial parallelism provided in studying several realizations
of random disorder in spin glasses or random-field systems
\cite{fang:14,baity-jesi:14,lulli:15,navarro:16}, or by parallelizing the energy
calculation in a system with long-range interactions \cite{gross:11}, for example. If
such additional parallelism is exploited, parallel tempering simulations on GPU show
excellent performance. Due to the almost embarrassingly parallel nature of the
algorithm, the scaling with the number of replicas is found to be almost ideal
\cite{gross:11}. For systems with only a few states such as the Ising model, it is
also possible to code several of the parallel tempering replicas in one machine word
yielding additional speedups. This will be discussed further in
Sec.~\ref{sec:disordered}.

\index{parallel~tempering|)}
\index{multicanonical~simulation|(}

\subsection{Multicanonical simulations}

While in parallel tempering barriers in the energy landscape are overcome by escaping
to high temperatures, in multicanonical simulations such regions of low probability
are artificially enhanced in an attempt to avoid a trapping of the system in
metastable states \cite{berg:92b}. To this end, one replaces the Boltzmann weight
proportional to $\exp(-\beta E)$ by a general weight function $W(E)$. As a result,
while the canonical energy distribution is
\[
P_{\beta}(E)=\frac{1}{Z_\beta}\Omega(E)e^{-\beta E},
\]
where $\Omega(E)$ is the density-of-states, and $Z_\beta$ denotes the canonical
partition function, the energy distribution in this modified ensemble then is
\begin{equation}
  P_\mathrm{muca}(E)=\frac{1}{Z_\mathrm{muca}}\Omega(E)W(E),
  \label{eq:muca}
\end{equation}
where $Z_\mathrm{muca}$ is the corresponding multicanonical partition function. As is
clear from Eq.~(\ref{eq:muca}), a flat distribution in energy is achieved for $W(E)
\propto \Omega^{-1}(E)$. Such weights can be determined iteratively by estimating the
density-of-states $\Omega(E)$ from a given simulation and adapting the weight
function accordingly, i.e.,
\begin{equation}
  \label{eq:weight_modification}
  W^{(n+1)}(E) \equiv \hat{\Omega}^{-1,(n)}(E) = W^{(n)}(E)/H^{(n)}(E),
\end{equation}
where $H^{(n)}(E)$ denotes the energy histogram in the simulation with weight
function $W^{(n)}(E)$. More elaborate weight iteration schemes that use the
accumulated information from all previous iterations can be devised
\cite{berg:96,janke:03a}, but we will not discuss these here. After the weights are
sufficiently converged to yield an approximately flat energy histogram, a production
run in the fixed ensemble with weights $W^{(\ast)}(E)$ is used to estimate the
observables of interest. As one of the paradigmatic applications, this scheme allows
one to study the strongly suppressed coexistence region for a
\index{phase~transition} first-order transition and determine the interface tension
between the phases \cite{janke:92}. Generalizations to reaction coordinates other
than the energy, for example to magnetizations \cite{berg:93} or bond and cluster
numbers \cite{wj:95a,weigel:10d} are also possible.

Although the problem of simulations with the general weights $W(E)$ appears to be
very similar to the special case $W(E) = \exp(-\beta E)$ of the canonical ensemble,
and hence the methods of parallelization outlined in Sec.~\ref{sec:canonical} should
be applicable, there is an important difference: while in the canonical case the
ratio $W(E')/W(E) = \exp(-\beta \Delta E)$ entering the Metropolis acceptance
criterion (and similar expressions for the heatbath method) depends on energy only
through the difference $\Delta E = E'-E$ incurred by the present move, the dependence
in the generalized-ensemble case is on $E'$ and $E$ individually. As a result, the
acceptance probability for each move depends directly on the total value of $E$, and
it becomes impossible to use a domain decomposition to flip spins in different
regions of the lattice in parallel. All transitions in the Markov chain hence must
occur in sequence. To still make use of parallel resources for simulations of this
kind, two complementary strategies have been proposed. The first approach relies on a
sub-division of the reaction-coordinate space (i.e., the energy for the simplest
case) into possibly overlapping intervals such that separate simulations run in
parallel can be used to cover all windows. The second strategy consists of setting up
independent Markov chains each of which covers all of the reaction-coordinate space,
but that communicate weight updates at certain intervals.

The windowing method was used in Ref.~[\refcite{weigel:11}]. In this approach, the
full energy range $[E_\mathrm{min},E_\mathrm{max}]$ is divided into $p$ windows
$[E_{i,\mathrm{min}},E_{i,\mathrm{max}}]$ and $p$ independent simulations with a
weight function $W^{(n)}(E)$ are used to simulate the system each with energies in
the corresponding window. For the individual simulations, it is crucial that the
current state is counted again if an attempted move leading the system outside the
energy window was rejected \cite{schulz:03}. As the initial choice
$W^{(0)}(E)=\mathrm{const}.$ can be used. Some scheme needs to be devised to create
appropriate initial configurations with energies inside the corresponding window. The
results of all simulations are then used to determine an estimate of the
density-of-states and hence an updated weight function $W^{(n+1)}(E)$ according to
Eq.~(\ref{eq:weight_modification}). Since the normalizations of the energy
distributions (\ref{eq:muca}) are different in each energy window, one needs to match
neighboring histograms at one or several common energy values at the window
boundary. Hence the method only works if the windows overlap by at least one energy
state. A GPU implementation for the Ising model was discussed in
Ref.~[\refcite{weigel:11}], where it was demonstrated that the density-of-states for,
e.g., a $64 \times 64$ system can reliably be estimated from energy windows as small
as $\Delta E = 16$ without systematic biases and with speed-up factors exceeding 100
as compared to the corresponding scalar CPU implementation.

\begin{figure}[tb]
  \centering
  \includegraphics[clip=true,keepaspectratio=true,width=0.75\columnwidth]{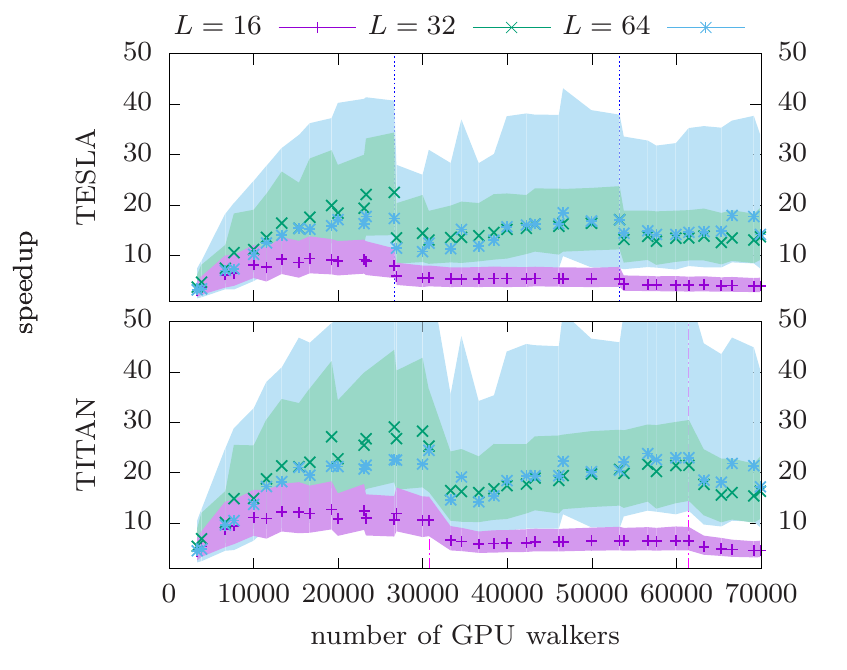}
  \caption{\label{fig:cudamuca}%
    Estimated speedup, i.e., reduction in wall-clock time required until convergence,
    for the parallel multicanonical method applied to the 2D Ising model on different
    GPUs (Tesla K20m, Titan Black) as a function of the number of walkers. The
    speedup is obtained compared to reference times from simulations on one CPU node
    equipped with two Intel E5-2640 6-core CPUs, using a total of 24
    hyper-threads. Data points mark the median speedup and the shaded areas indicate
    the confidence interval covering $2/3$ of the data.  }
\end{figure}

The second approach for parallel multicanonical simulations was proposed in
Ref.~[\refcite{zierenberg:13}] and uses $p$ parallel walkers that are unrestricted in
energy (or another reaction coordinate if that is chosen for the scheme) and work
with the same weight function $W^{(n)}(E)$. At the end of an iteration, the
histograms of individual runs are added up,
\[
  H^{(n)}(E) = \sum_{i=1}^p H^{(n)}_{i}(E),
\]
and the total histogram is used to determine the updated weight function
$W^{(n+1)}(E)$ according to Eq.~(\ref{eq:weight_modification}). This is then again
distributed to the $p$ walkers to perform runs of the next iteration
\cite{zierenberg:13}. The final production run can be performed in the same way,
pooling the final results from $p$ simulations to improve statistics. A GPU
implementation of this method was recently introduced in
Ref.~[\refcite{gross:17}]. It parallelizes only over independent walkers, thus
resulting in particularly simple and easily adaptable code. In the example
implementation for the 2D Ising model, it uses random-site selection for the
individual spin updates, mainly in order to be able to scale the number of updates in
units of individual spin flips instead of in units of sweeps. In order to achieve
good coalescence of memory accesses, the same random-number sequence is used to
select the spins in all replicas, but different, uncorrelated sequences are used to
decide about the acceptance of spin flips. Instead of collecting individual
histograms $H^{(n)}_{i}(E)$ to be added up at the end of each iteration, it turns out
to be more efficient for each thread to add events to a unique histogram $H^{(n)}(E)$
kept in global memory using \index{atomic~operation} atomic operations for the
increments. Since the number of possible energy values typically grows faster than
the number $p$ of walkers, this leads to excellent performance as collisions in
accessing histogram entries are rare. For the total times per spin-flip, including
the time spent on histogram and weight updates, we arrive at peak performances of
0.22 ns and 0.16 ns for the Tesla K20m and GTX Titan Black cards, respectively, which
corresponds to a 15--21 times speedup as compared to the performance of an MPI code
on a full dual-CPU node with a total of 12 cores (24 hyper-threads) with Intel Xeon
E5-2640 CPUs \cite{gross:17}. This optimal performance is found for fully loading the
GPUs with threads, i.e., for the maximum \index{occupancy} occupancy, corresponding
to $30\,720$ threads for the Titan Black and $26\,624$ threads for K20m. The total
speedup of the parallel implementation also depends on the effect of the parallel
calculation on the number of required iterations until divergence, which is found to
be slowly decreasing with $p$ \cite{gross:17}, at least if a number of equilibration
updates in between iterations ensures that the walkers are thermalized with respect
to the updated weights before collecting statistics for the next iteration. The total
speedup in the time-to-solution for the parallel multicanonical code on GPU as a
function of $p$ is shown in Fig.~\ref{fig:cudamuca}. The characteristic shape of
these curves is related to occupancy effects of the devices with the vertical lines
indicating the optimal number of threads mentioned above. The optimal speedups in the
time-to-solution are about 20 for the Tesla K20m and about 25 for the Titan Black, so
even a bit larger than the hardware speedups. This effect is attributed to the fact
that the histograms collected by independent walkers are somewhat less correlated
than those sampled by a single simulation for the same total number of hits. The
observed scaling properties are quite good, although it is clear that the necessary
equilibration steps will asymptotically destroy parallel scaling as the total
parallel work is of the form $W=W_0+pT$, where $W_0$ denotes the sampling sweeps and
$T$ the equilibration steps, such that the work $W/p$ per walker approaches a
constant as $p\to\infty$.\footnote{We note that the number $W_0$ of samples until
  convergence will itself (weakly) depend on $p$ \cite{gross:17}, but this does not
  affect the argument for a diminishing scaling efficiency as $p\to\infty$.}

\index{multicanonical~simulation|)}
\index{Wang-Landau~sampling|(}

\subsection{Wang-Landau update}

The Wang-Landau method \cite{wang:01a,wang:01b} can be seen as a different technique
for determining the optimal weights in a multicanonical simulation or, alternatively,
as a method for directly estimating the density-of-states. It is not a traditional
Markov-chain method as it changes the ensemble at each step, but a variant has been
classified as a stochastic approximation algorithm \cite{liang:06}. The algorithm
continuously modifies a working estimate $g(E)$ (initialized as $g(E)=1$ $\forall E$)
for the density-of-states $\Omega(E)$ by multiplying it for the currently visited
energy bin by a modification factor $f$, initially chosen to be $f=e$. Spin flips for
the transition $E\to E'$ are accepted with probability
\[
p_\mathrm{acc} = \min[1,g(E)/g(E')].
\]
If an energy histogram $H(E)$ recorded during the updating procedure is
``sufficiently flat'' \cite{wang:01a}, the modification factor is reduced as
$f\to \sqrt{f}$ and $H(E)$ is reset to zero. In practice, flatness is often declared
if the minimum histogram bin has at least 80\% of the average number of hits. The
algorithm terminates once $f$ has reached a certain accuracy threshold, for instance
$10^{-8}$. While the method appears to converge well in general, it has been noted
that the accuracy cannot be arbitrarily increased with the given schedule of reducing
$f$ and there is actually a residual error that does not diminish for longer
simulations \cite{qiliang:03}. A number of schemes have been suggested to correct
this, in particular a $1/t$ decay of the modification factor after an initial
exploration phase \cite{belardinelli:07} and a class of methods dubbed stochastic
approximation Monte Carlo \cite{liang:06,liang:07}.

Parallelization of the algorithm proceeds along similar lines as for the
multicanonical method. The energy range can be divided into windows\footnote{Note
  that the original proposal in Ref.~[\refcite{wang:01b}] contained a mistake in that
  after the rejection of a move that would have led the system outside of the energy
  window the current configuration was not counted again. This was later on corrected
  in Refs.~[\refcite{schulz:02,schulz:03}].} that are sampled by individual walkers
\cite{wang:01b}. Window overlaps are then required to allow for a matching together
of the pieces of the density-of-states. A GPU implementation of this scheme was
discussed in Ref.~[\refcite{weigel:11}]. Some inefficiencies often occur in such
schemes due to the random nature of the run-time of the algorithm for the different
windows, and additional load balancing would be required to alleviate this effect. An
alternative approach is somewhat similar in spirit to the parallel multicanonical
method of Ref.~[\refcite{zierenberg:13}] in that it employs a large number of
parallel walkers \cite{yin:12}. As the continuous weight modification after each flip
would serialize all updates, however, this condition is relaxed and the walkers work
with separate estimates $g_p(E)$ that are only synchronized at certain intervals. A
more flexible approach combines aspects of parallel tempering with the Wang-Landau
method \cite{vogel:13}, such that Wang-Landau simulations are performed in
overlapping energy windows, and replica-exchange moves are attempted between walkers
in neighboring windows. Several independent walkers can be employed in each window,
for which the flatness of histograms is assessed separately; their estimates are
averaged before changing the modification factor and moving to the next
iteration. Improved load balancing is attempted by choosing energy windows of uneven
size, such that the expected convergence time between intervals stays the same
\cite{vogel:13}.

\index{Wang-Landau~sampling|)}
\index{population~annealing|(}

\subsection{Population annealing}
\label{sec:PA}

A more recent addition to the arsenal of generalized-ensemble simulations is not
drawn from the class of Markov chain Monte Carlo algorithms, but instead hails from
the kingdom of sequential Monte Carlo methods \cite{doucet:13}. Population annealing
was first suggested in Refs.~[\refcite{iba:01,hukushima:03}] and more recently
rediscovered and improved in Ref.~[\refcite{machta:10a}]. In this approach, a large
population of replicas of the system are simulated at the same temperature. At
periodic intervals, the temperature is lowered and configurations are resampled
according to their relative Boltzmann weight at the lower temperature. This process
is continued until a pre-defined target temperature has been reached. Measurements of
observables are then taken as population averages at a given temperature. The
algorithm can be summarized as follows:
\begin{enumerate}
\item Set up an equilibrium ensemble of $R_0 = R$ independent replicas of the system
  at inverse temperature $\beta_0 = 0$.
\item To create an approximately equilibrated population at $\beta_i > \beta_{i-1}$,
  resample configurations $j = 1,\ldots, R_{i-1}$ with their relative Boltzmann
  weight $\tau_i(E_j) = \exp[-(\beta_i-\beta_{i-1})E_j]/Q_i$, where
  \begin{equation}
    Q_i \equiv Q(\beta_{i-1},\beta_i) = \frac{1}{R_{i-1}}
    \sum_{j=1}^{R_{i-1}} \exp[-(\beta_i-\beta_{i-1})E_j].
    \label{eq:Q}
  \end{equation}
\item Update each replica by $\theta$ sweeps of a Markov chain Monte Carlo (MCMC)
  algorithm at inverse temperature $\beta_i$.
\item Calculate estimates for observable quantities ${\cal O}$ as population averages
  $\sum_{j=1}^{R_i} {\cal O}_j/R_i$.
\item If the the target temperature $\beta_\mathrm{f}$ has not been reached, goto
  step (ii).
\end{enumerate}
The approach is similar in spirit to parallel tempering, but it is intrinsically much
more suitable for parallel computing due to the large populations required (typically
at least $10^4$ replicas, often also \cite{wang:15a,wang:15b} $10^6$ or
$10^7$). Also, it allows access to certain population-related quantities such as the
free energy, which are more difficult to measure in parallel tempering (but see
Ref.~[\refcite{wang:15c}]). A number of improvements, such as adaptive temperature
steps, time steps and population sizes have been proposed
\cite{barash:16,weigel:17a}. Also, it is possible to use a multi-histogram analysis
to improve the statistical quality of data and provide results for any temperature
point \cite{barash:16} as well as weighted averages of simulations with smaller
population sizes to reduce population-size related bias
\cite{machta:10a,wang:15a}. As it stands, the approach is not directly efficient for
simulating first-order transitions \cite{barash:17}.

An implementation of population annealing for GPU was discussed in
Ref.~[\refcite{barash:16}], using as the example application a 2D Ising model. The
parallelization of the spin updates can rely on the same approaches as for the
canonical simulations discussed in Sec.~\ref{sec:canonical}. Replica-level
parallelism, where the threads of a block update the same locations in different
replica, is the inherent parallelism of the method and it ensures that population
annealing can be efficiently implemented on GPU independent of the model. For the
application of the method to the Ising model, this approach is combined with
spin-level parallelism, using \index{domain~decomposition!checkerboard} tiling and a
checkerboard update as in the implementation discussed in Sec.~\ref{sec:checkerboard}
above. The Philox generator is used for the Metropolis update, cf.\
Sec.~\ref{sec:RNG}. The resampling is implemented by calculating the weight functions
$Q_i$ of Eq.~(\ref{eq:Q}) in parallel and implementing the scan pattern
\index{algorithmic~pattern!scan} (cf. Sec.~\ref{sec:patterns}) to determine the new
position of copies of replicas in the resampled population. Measurements as averages
over the population are computed using parallel \index{algorithmic~pattern!reduction}
reductions, using \index{atomic~operation} atomic operations to combine the partial
results from different thread blocks. In total, this approach yields excellent
speedups as compared to a serial implementation, the peak performance on a Tesla K80
GPU being around 230 times faster than the serial code on an Intel Xeon E5-2683 v4
CPU. The overhead for resampling is found to be rather small for typical values
$\theta$ of the number of rounds of spin flips performed in between resampling steps,
coming in at less than 15\% of the runtime for $\theta \ge 10$
\cite{barash:16}. Multi-spin coding can be used to increase the peak performance to
less than 10 ps per spin flip on the K80, resulting in speedups of a factor of 2400
against the serial CPU code. In this setup, the compression of 32 or 64 spins into a
single word leads to a significant relief of bandwidth pressure for memory transfers,
such that calculations are then typically limited by their arithmetic density. The
cost of generating random numbers therefore turns into an important consideration. If
the underlying, high-quality RNG (in this case Philox \cite{salmon:11}) is used for
generating the random numbers for all multi-spin coded copies, the overall
performance benefit of multi-spin coding is very moderate \cite{barash:16}. As an
alternative, a combination with a cheap linear-congruential generator dealing with
the copies coded together that is re-seeded for each sweep by Philox provides good
statistical quality and the excellent performance results quoted above
\cite{barash:16}.

\index{population~annealing|)}
\index{quenched~disorder|(}

\section{Disordered systems}
\label{sec:disordered}

Simulations of systems with quenched disorder, in particular spin glasses and
random-field systems \cite{young:book}, are extremely demanding computationally. The
reason is twofold: firstly the rugged free-energy landscape of such systems results
in extremely slow relaxation, and secondly the necessary disorder average means that
all calculations need to be repeated for many thousand disorder realizations. The
relaxation can be sped up with the help of the generalized-ensemble methods discussed
above, but they only provide a moderate improvement as still typically the relaxation
times increase very steeply with system size. Parallel tempering is the technique
most commonly used for Monte Carlo simulations of such problems
\cite{katzgraber:06a,banos:10}, but also multicanonical and Wang-Landau methods are
employed \cite{berg:98a,malakis:06} and, more recently, population annealing
simulations \cite{wang:14,wang:15b}.

This situation turns such systems into ideal targets for massively parallel
simulation methods. While the required generalized-ensemble simulation techniques are
more or less well suitable for parallel computing as outlined in
Sec.~\ref{sec:generalized}, the average over disorder provides an additional
dimension along which simulations are trivially parallel, and so ideal scaling can be
expected. For spin glasses a number of independent initiatives have provided
implementations of simulation codes for massively parallel hardware.  The main focus
has been on systems with discrete spins such as Ising and Potts spin glasses, where
the Ising case corresponds to the Hamiltonian (\ref{eq:ising_model}) with couplings
$J_{ij}$ typically drawn from a Gaussian or a bimodal distribution. A number of
different implementations on GPU each use some mixture of the same general
ingredients \cite{weigel:10a,fang:14,lulli:15}: \index{parallel~tempering} parallel
tempering, \index{domain~decomposition!checkerboard} checkerboard updates and tiling,
multi-spin coding \index{multi-spin~coding} across different disorder
realizations,\footnote{We note that for spin-glass systems often the same random
  number is used to decide about the flipping of all spins coded together as it is
  argued that the randomness in disorder realizations together with the properties of
  bond chaos \cite{bray:87} lead to a sufficient decorrelation of individual
  trajectories \cite{lulli:15}.} carefully tailored setups for random-number
generation. Some attention has also been paid to systems with continuous degrees of
freedom, in particular the Heisenberg spin glass, where the exceptional
floating-point performance \index{floating-point~performance} of current GPUs can be
brought to the fore, in particular if single-precision variables are used for the
spins and the special-function units can be employed
\cite{yavorskii:12,baity-jesi:14}. A different architecture has been used by the
JANUS initiative that has constructed special-purpose machines for simulations of
discrete-spin glass models based on field-programmable gate arrays \index{FPGA}
(FPGAs) \cite{belleti:09,baity-jesi:14a}. Comparing these to GPU implementations, the
overall throughput is similar, but the FPGA machines allow to simulate single
realizations at higher spin-flip rates than the GPU codes can
\cite{fang:14,lulli:15}. Massively parallel implementations of codes for random-field
systems have also focused on the Ising case, corresponding to the Hamiltonian
(\ref{eq:ising_model}) with local random fields $h_i$ but uniform couplings
$J_{ij} = J$. There has been significantly less work on such problems using massively
parallel machines, but efficient implementations can be achieved using techniques
almost identical to those for the spin glasses, a recent example is provided in
Ref.~[\refcite{navarro:16}].

\index{quenched~disorder|)}

\section{Summary}
\label{sec:conclusions}

The aim of this chapter was to give a general introduction to aspects of (massively)
parallel computing relevant for practitioners in the field of computer simulations in
statistical physics. We provided some general background including the scaling theory
of parallel performance, an overview of the available parallel hardware with a focus
on graphics processing units, and an outline of the most important algorithmic
skeletons or patterns in parallel computing. In the application part we went on to
discuss the parallelization of canonical Monte Carlo algorithms such as single-spin
flip simulations of lattice models, but also the cluster updates that allow to tackle
the slowing down of dynamics close to critical points. After giving an outline of the
specific problems of random-number generation for massively parallel applications and
their solution, we went on to discuss the challenges posed by parallel
implementations of generalized-ensemble simulation algorithms used for simulating
systems with complex free-energy landscapes, including parallel tempering,
multicanonical and Wang-Landau simulations as well as the population annealing
method. Most of the tricks of this trade come together in the simulation of
disordered systems with massively parallel resources, an application which seems to
be an ideal fit for devices such as GPUs, FPGAs and Xeon Phi.

\section*{Acknowledgments}

I thank Yurij Holovatch for his kind invitation to present one of the Ising
Lectures at the Institute for Condensed Matter Physics of the National
Academy of Sciences of Ukraine, Lviv, Ukraine.

I gratefully acknowledge the contributions to the work reviewed here by my
collaborators, in particular Lev Barash, Michal Borovsk\'{y}, Jonathan Gross,
Alexander Hartmann, Wolfhard Janke, Jeffrey Kelling, Ravinder Kumar, Markus Manssen,
Lev Shchur, Taras Yavors'kii, and Johannes Zierenberg. I also thank Axel Arnold,
Benjamin Block, Eren El\c{c}i, Ezequiel Ferrero, Nikolaos Fytas, Helmut Katzgraber,
Ralph Kenna, David Landau, Jonathan Machta, and Peter Virnau for many useful
discussions relating to the subject.

This work was partially supported by the European Commission through the IRSES
network DIONICOS under Contract No.\ PIRSES-GA-2013-612707, by the German Research
Foundation (DFG) in the Emmy Noether Program under contract No.\ WE4425/1-1, and
through the Royal Society under contract No.\ RG140201.

%\bibliographystyle{ws-rv-van}
%\bibliography{citeulike_nourl_noissn}
%\bibliography{citeulike_nodoi_noissn}

%\blankpage
%\printindex[aindx]                 % to print author index
\printindex                         % to print subject index

\end{document}